\documentclass[journal,twocolumn]{IEEEtran}
\usepackage{cite}
\usepackage[english]{babel}

\usepackage{float}
\usepackage{amsthm}
\usepackage{amsmath}
\usepackage{commath}
\usepackage{nccmath}
\usepackage{empheq}
\usepackage{etoolbox}
\usepackage{multirow}
\usepackage{graphicx}
\usepackage{tabularx}
\usepackage{subcaption}
\usepackage[colorlinks=true, allcolors=blue]{hyperref}
\usepackage{algorithmicx}
\usepackage{array}
\usepackage{comment}
\usepackage{relsize,amsfonts}
\usepackage[ruled,lined,linesnumbered]{algorithm2e}

\newcolumntype{C}{>{\centering\arraybackslash}X}

\DeclareSymbolFont{myletters}{OML}{ztmcm}{m}{it}
\DeclareMathSymbol{\uplambda}{\mathord}{myletters}{"15}
\usepackage{filecontents,lipsum}
\DeclareMathOperator*{\argmin}{argmin}

\theoremstyle{definition}
\newtheorem{defn}{Definition}
\newtheorem{prop}{Proposition}
\newtheorem{lem}{Lemma}
\newtheorem{rem}{Remark}
\newtheorem{thm}{Theorem}
\newtheorem{Assume}{Assumption}

\newcommand{\zerodisplayskips}{%
  \setlength{\abovedisplayskip}{0pt}%
  \setlength{\belowdisplayskip}{0pt}%
  \setlength{\abovedisplayshortskip}{0pt}%
  \setlength{\belowdisplayshortskip}{0pt}}
\appto{\normalsize}{\zerodisplayskips}
\appto{\small}{\zerodisplayskips}
\appto{\footnotesize}{\zerodisplayskips}

\title{Rethinking Clustered Federated Learning in NOMA Enhanced Wireless Networks}

\author{Yushen Lin,~\IEEEmembership{Student Member,~IEEE,}
        {Kaidi Wang,}~\IEEEmembership{Member,~IEEE,}
        {and Zhiguo Ding,~\IEEEmembership{Fellow,~IEEE}}
\vspace{-2em} 
\thanks{Y. Lin and K. Wang are with the School of Electrical and Electronic Engineering, The University of Manchester, M13 9PL, U.K. (e-mails: yushen.lin@student.manchester.ac.uk and kaidi.wang@ieee.org).

Zhiguo Ding is with the Department of Electrical Engineering and Computer Science, Khalifa University, Abu Dhabi, UAE, and the Department of Electrical and Electronic Engineering,
University of Manchester, Manchester, UK (e-mail: zhiguo.ding@manchester.ac.uk).

Part of this work has been submitted to IEEE 99-th Vehicular Technology Conference: VTC2024-Spring \cite{VTC_Lin}.}
}

\begin{document}
\maketitle

\begin{abstract}
This study explores the benefits of integrating the novel clustered federated learning (CFL) approach with non-orthogonal multiple access (NOMA) under non-independent and identically distributed (non-IID) datasets, where multiple devices participate in the aggregation with time limitations and a finite number of sub-channels. 
A detailed theoretical analysis of the generalization gap that measures the degree of non-IID in the data distribution is presented. Following that, solutions to address the challenges posed by non-IID conditions are proposed with the analysis of the properties. Specifically, users' data distributions are parameterized as concentration parameters and grouped using spectral clustering, with Dirichlet distribution serving as the prior. The investigation into the generalization gap and convergence rate guides the design of sub-channel assignments through the matching-based algorithm, and the power allocation is achieved by Karush–Kuhn–Tucker (KKT) conditions with the derived closed-form solution. The extensive simulation results show that the proposed cluster-based FL framework can outperform FL baselines in terms of both test accuracy and convergence rate. Moreover, jointly optimizing sub-channel and power allocation in NOMA-enhanced networks can lead to a significant improvement.

\end{abstract}



\maketitle
\vspace{-0.1cm}
\section{Introduction}
As stated in \cite{Cisco}, there will be more than 29 billion networked devices by 2023. In light of the rapid growth of mobile applications and networked devices, traditional machine learning frameworks face new challenges in supporting the vast amount of data \cite{Y_Liu_FL_literature}. In this context, federated learning (FL) attracts the great attention of researchers \cite{FL_NOMA_6G}. FL offers better privacy and efficiency than traditional learning schemes by periodically sharing the local model updates with the server. It is widely acknowledged that the increased size of data in the training process can enhance the performance of the model. However, the constraints imposed by wireless networks, i.e., a limited number of sub-channels and stringent energy consumption requirements, can adversely impact the performance of FL due to reduced user participation. Moreover, this problem becomes even more severe when dealing with non-independent and identically distributed (non-IID) datasets \cite{FL_Wireless_Qin}. Consequently, non-orthogonal multiple access (NOMA) is gaining attraction as a viable solution for its ability to boost spectrum efficiency and support massive connectivity \cite{NOMA_user_association, NOMA_MEC_Ding}. Due to the above characteristics, employing NOMA to improve the performance of FL is well-motivated. Specifically, greater cooperation and engagement across the spectrum can be achieved by involving more devices in the training process. Therefore, it is crucial to develop advanced FL frameworks and design wireless networks accordingly.
\vspace{-0.1cm}
\subsection{Prior Works}
Since Google's pioneering study on FL \cite{FL_original}, there has been a surge of research interest in FL and its design in wireless networks. In the literature, the studies have been investigated from the following two perspectives. Firstly, the state-of-the-art FL and Clustered FL (CFL) frameworks are investigated. Then, the recent research related to wireless communication design for FL is reviewed.

\subsubsection{Prior Works on FL and CFL}
In real-world scenarios, i.e., cross-device or cross-silo, the data distribution tends to be non-IID across the devices \cite{FL_literature_1}. Training on such distribution can lead to weight divergence and client drift, which consequently hinder the model generalization as noted in several works \cite{FL_with_non_iid,scaffold_FL}. To mitigate the adverse effects caused by the non-IID dataset, extensive studies have been investigated \cite{CFL_model_agnostic, HypCluster, FL_HC, FedMe}. In \cite{CFL_model_agnostic}, the authors present a CFL framework to address the issue of the divergence in the data distributions of local devices, where the clustering is based on the cosine similarity between the weight updates of different devices. The framework in \cite{HypCluster} assigns devices to the cluster model that yields the lowest loss on the device's local data, while FL+HC \cite{FL_HC} utilizes hierarchical clustering (HC) by treating devices' gradients as features on various distance metrics. In \cite{FedMe}, the authors apply K-means to devices’ model weights to address the problem caused by non-IID datasets. The ultimate goal of these methods is to group devices' model parameters or gradients, either through a one-shot or iterative approach. 


\subsubsection{FL in Wireless Networks}
There are extensive studies that focus on FL in wireless networks \cite{kaidi2023fl2, FL_fading_channels, joint_communications_FL, kaidi_AOI_FL, Contribution_based_WFL, Latency_constrained_WFL, FL_nanyang}. In \cite{kaidi2023fl2}, Wang et al. investigate the FL system over networks with non-IID datasets, focusing on minimizing age-of-information (AoI) as a metric. The study proposes an AoI-based device selection scheme to address weight divergence and convergence issues in non-IID datasets. In \cite{FL_fading_channels}, the authors propose techniques to implement distributed stochastic gradient descent over a shared noisy wireless channel. Chen et al.\cite{joint_communications_FL} tackle the issue of training FL algorithms over wireless networks, taking into account wireless constraints such as packet errors and resource availability, where the Hungarian algorithm is used to select devices based on data size. In \cite{kaidi_AOI_FL}, the authors 
investigate the concept of the AoI into FL over wireless networks, aiming to optimize various aspects for achieving better performance, i.e., device selection and resource allocation. In \cite{Contribution_based_WFL}, a min-max optimization problem is formulated into a primal-dual setup. The modified Truncated Monte Carlo method is employed for estimating device contributions, leading to improvements in model generalization, convergence, and device-level performance in FL. In \cite{Latency_constrained_WFL}, a study of maximizing model accuracy within a given total training time budget in latency-constrained FL in wireless networks is investigated. A lower bound on the training loss is derived to guide the design of the solutions. The paper \cite{FL_nanyang} addresses the challenges of deploying FL in wireless networks by proposing the Adaptive Quantization Based on Ensemble Distillation (AQeD) scheme, which efficiently manages the complexities arising from model heterogeneity and communication costs.

NOMA, as one of the promising technologies for the next generation of wireless networks  \cite{NOMA_survey_ding, NOMA_survey_fang, ZhangSWIPT, Lin_unsupervised, Wenqi_NOMA, Qin_NOMA}, is increasingly considered as a potential solution for FL \cite{Adaptive_FL_NOMA, NOMA_FL_cost, wanli_FL, RIS_FL_NOMA, Scheduling_allocation_FL, Privacy_NOMA_FL}.
Specifically, aiming to reduce communication latency without sacrificing test accuracy, NOMA in FL model updates is investigated in \cite{Adaptive_FL_NOMA}.
 It also takes into account capacity-limited channels and validates the proposed adaptive compression scheme using multiple datasets. In \cite{NOMA_FL_cost}, the paper presents a novel approach to optimize FL in wireless networks through a NOMA-assisted FL system, leveraging a layered algorithm to address a non-convex optimization problem, and demonstrating significant improvements in system cost and efficiency. In \cite{wanli_FL}, it presents a novel framework that integrates NOMA and over-the-air federated learning (AirFL) through simultaneous transmission and reflection using a reconfigurable intelligent surface (STAR-RIS). In \cite{RIS_FL_NOMA}, for further improving spectrum efficiency, the
authors investigate the use of mobile 
reconfigurable intelligent surface (RIS) in NOMA networks where the considered system is optimized through Deep Deterministic
Policy Gradient and FL. For \cite{Scheduling_allocation_FL}, the research focuses on the scheduling policies and power allocation strategies for FL in a NOMA-based mobile edge computing environment, to maximize the weighted sum data rate. Liu et al. \cite{Privacy_NOMA_FL} explore the potential for enhancing privacy in FL wireless networks using uncoded transmission strategies coupled with adaptive power control. Their research illustrates the advantages of implementing NOMA over the orthogonal multiple access (OMA) scheme, showing the improvements in both privacy and learning performance.

\vspace{-0.1cm}
\subsection{Motivation and Contributions}
Despite the numerous and critical directions of FL in wireless networks that have been investigated in the existing papers, the study of NOMA-enhanced FL design still remains a notable gap. Specifically, there is a lack of joint optimization of the NOMA transmission in relation to sub-channel assignment and power allocation, along with the training accuracy of the FL, especially when dealing with non-IID data distribution. 
Moreover, advanced and efficient FL frameworks are not adequately explored in the context of FL in wireless networks, as training on non-IID data can further negatively impact FL performance.
Since the data is stored locally, it is impractical to transfer all the data to the server for clustering due to high communication costs and privacy constraints. Most existing CFL frameworks demand high computational resources or extensive data transmission to the server, which may not be practical given the computational and transmission overhead in wireless networks. For example, the similarities between each user's gradient are required to be computed each round, or all users need to be engaged in the first round for the clustering purpose.
To address these challenges, there is a pressing need to develop a mathematically tractable FL in wireless networks framework tailored for non-IID data distributions. In this light, a novel NOMA-assisted CFL framework is proposed. The contributions are elaborated as follows:

\begin{enumerate}
\item  The factors affecting the generalization gap and convergence rate are investigated. Based on the theoretical insights, a new CFL-based framework empowered by NOMA is explored, where a subset of devices join the aggregation in each communication cycle through the NOMA scheme. The proposed framework does not require additional calculations on the user devices nor does it hard modify the established FL protocols, in particular does not demand a huge communication cost between devices and the server.

\item The server manages the calculation of the clustering method that is designed based on the degree of heterogeneity. Only limited information is requested from the device to reduce the computational burden on the devices. This is determined using the probability mass function of the Multinomial-Dirichlet ($\mathcal{MD}$) distribution and solved by the Broyden–Fletcher–Goldfarb–Shanno (BFGS) algorithm. Within this framework, parameters are categorized using spectral clustering, and the properties are analyzed. 

\item The resource allocation problem is decoupled into two parts, including the sub-channel assignment and power allocation problems. The first is tackled by the matching-based algorithm and the second is solved by the Karush–Kuhn–Tucker (KKT) conditions, for which a closed-form solution is derived. 

\item  The extensive simulation results show that the proposed schemes have superior performance and effectiveness over the FL baselines. This not only shows the inherent merits of the proposed design but also demonstrates how the proposed resource allocation and sub-channel assignment can dynamically enhance performance.  Moreover, it reduces the fluctuations during training and the impacts of fine-tuning on performance.

\end{enumerate}

\section{System Model}\label{section:system}

\begin{figure}[!htbp]  
    \centering
  \subfloat[More skewed distribution.\label{Fig:alpha_001}]{%
       \includegraphics[width=0.50\linewidth]{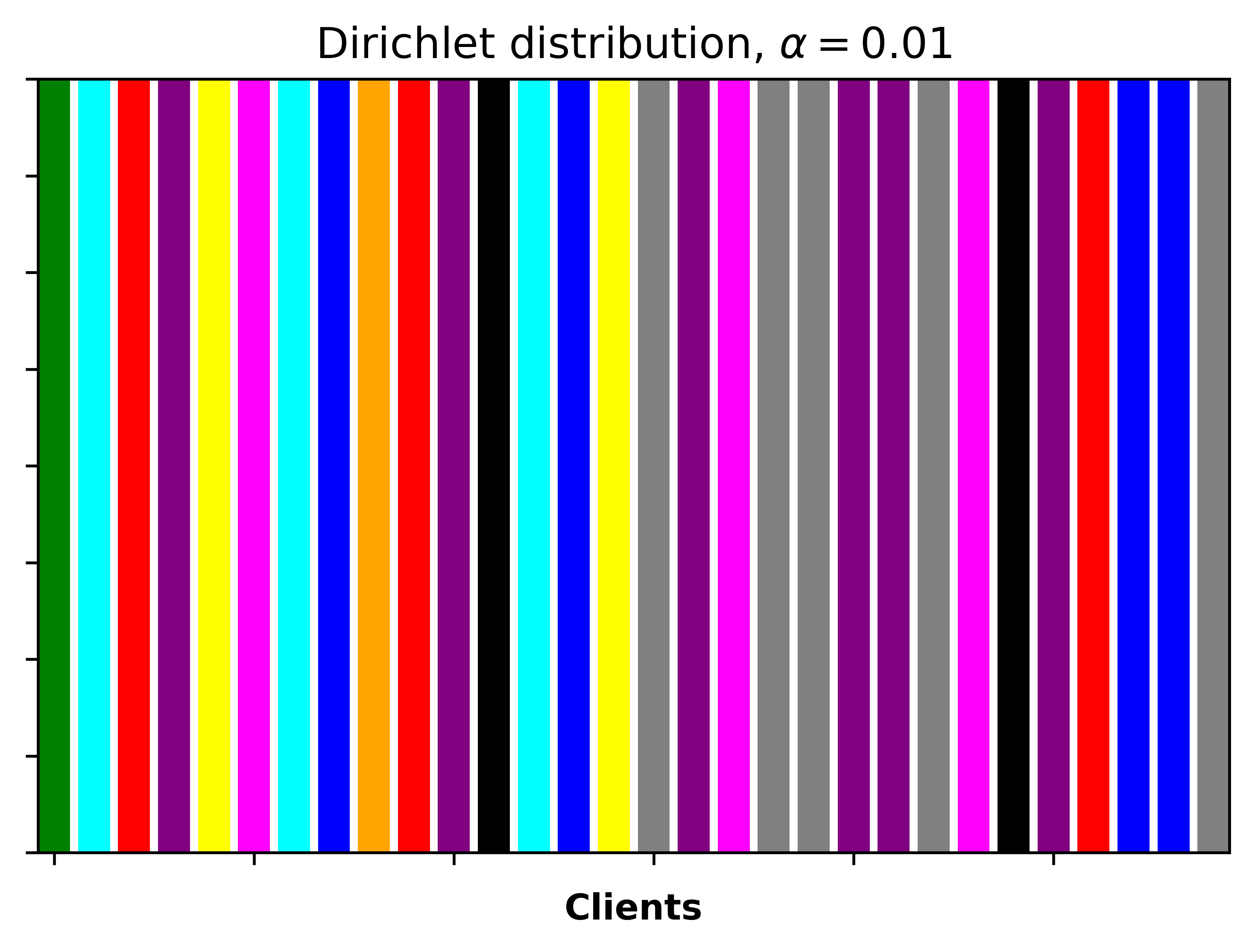}}
    \hfill
  \subfloat[More uniform distribution.\label{Fig:alpha_100}]{%
        \includegraphics[width=0.50\linewidth]{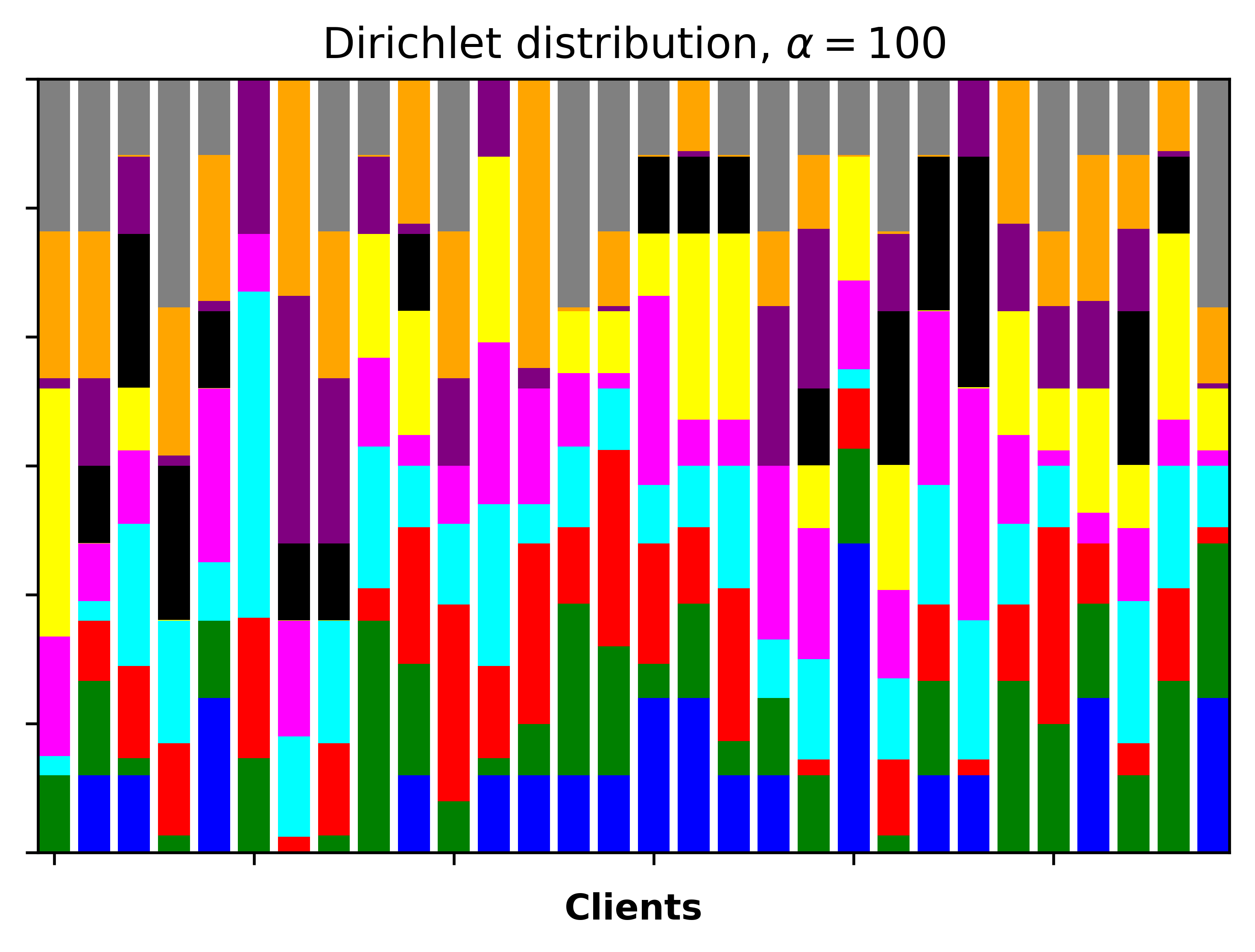}}
         \hfill
    \\
  \caption{An illustration of population distribution when $N = 30$, where the distribution among classes is represented with different colors, i.e., each user has a different number of data label(s) of data under different $\alpha$ as shown in (a) and (b).}
  \label{fig1:user_distribution} 
\end{figure}

In this section, a novel framework of CFL is designed and deployed in a NOMA-assisted wireless network. A server connects to $N$ users via $K$ sub-channels, where $N > K$. The collections of all users and sub-channels are denoted by $\mathcal{N} = \{1, 2, \cdots, N\}$ and $\mathcal{K} = \{1, 2, \cdots, K\}$, respectively. The subset of the device selected for aggregation and the maximum allowable time for training and offloading are denoted by $\boldsymbol{S}$ and $T_{\text{max}}$, respectively. The size of the local model of all devices is denoted by $D$. It is assumed that at most two users\footnote{Engaging more than two users can further enhance system throughput, as demonstrated in many existing studies\cite{hybrid_NOMA,NOMA_survey_ding}. However, this improvement comes at the cost of augmenting the processing complexity within the SIC receiver. In this work, the two-user NOMA case is considered to maintain a low-complexity system.} can be allocated to each sub-channel. Each device is equipped with a central processing unit (CPU) that is assumed to have different levels of computational capability, and $\beta_i$ is the number of $i$-th user's training samples. In order to resemble a practical scenario, the data distribution of the device is assumed to be non-IID. The local datasets are spawned from a unique distribution, specifically, $\mathcal{D}_i$ from Dirichlet distribution\footnote{
It is important to specify that when the data distribution of each user is mentioned, it means that the class labels for each user follow a multinomial distribution, which is a special case of a categorical distribution. Furthermore, the parameters of this multinomial distribution are given a conjugate prior drawn from the Dirichlet distribution.} within the population distribution, where $\mathcal{D}_i$ denotes the distribution associated with the $i$-th device drawn from the population distribution, $\mathcal{D}_i \neq \mathcal{D}_j$,  $\forall i\neq j$. The degree of data heterogeneity is parameterized by $\alpha$, where $\alpha \in \mathbb{R_{++}}$ denotes the concentration parameter in Dirichlet distribution. As $\alpha \rightarrow \infty$, local class labels become more uniform, and as $\alpha \rightarrow 0$, these labels become less uniform, as shown in Fig.~\ref{fig1:user_distribution}.

\vspace{-0.1cm}
 \subsection{Clustered Federated Learning Model}
To demonstrate the motivation of the proposed framework, a mathematical analysis of FL is first provided to investigate performance degradation caused by non-IID data. The mild assumptions are considered prior to introducing the rationale:
 \begin{Assume} \label{Assumption:Assumption_1}
    \textit{The underlying  distributions of each dataset $\tilde{\mathcal{D}}_i$, are drawn from $\mathcal{MD}\left(\alpha_i^*\right)$ distribution with concentration parameters $\boldsymbol{\alpha}_i^*=\left[\alpha_{i 1}^*, \ldots, \alpha_{i C}^*\right]$ under a reference distribution, where $C$ denotes the number of categories, and $ \mathcal{MD}(\boldsymbol{\alpha})$ is a $\mathcal{MD}$ distribution with concentration parameters $\boldsymbol{\alpha}$.}
 \end{Assume}

\begin{Assume} \label{assumption:bound_variance}
    For all clients $n$, the variance of the local gradient w.r.t. the global gradient is bounded by $G$ when there is no perturbation.
\end{Assume}
 
Denote $\mathcal{H}$ by hypothesis space on the input space, the empirical learned model by ${\widehat{\boldsymbol{w}}} = \argmin_{\boldsymbol{w}\in\mathcal{H}} F(\boldsymbol{w})$, and $\widehat{\boldsymbol{w}}^* = \argmin_{\boldsymbol{w}\in\mathcal{H}} F_u(\boldsymbol{w})$, where $F(\boldsymbol{w})$ and $F_u(\boldsymbol{w})$ respectively denote global loss function and the global loss function on the unseen data, i.e., data that the model has not encountered during its training phase. The weighted Rademacher complexity is denoted by $R_{\boldsymbol{\alpha}}(\mathcal{H})$.
\begin{thm}\label{thm:core_motivation}
     \textit{
    The generalization gap of the empirical learned global model on unseen data $F_u\left(\widehat{\boldsymbol{w}}\right) - F_u\left({\boldsymbol{\widehat{w}}^*}\right)$ follows a probability of at least $1 - \delta$ for any $\delta \in (0,1)$}:
    \begin{equation}
        F_u\left(\widehat{\boldsymbol{w}}\right) - F_u\left({\boldsymbol{\widehat{w}}^*}\right) \leq 2 R_{\boldsymbol{\alpha}}(\mathcal{H}) + \sqrt{\sum _{i \in \boldsymbol{S}}\frac{A^2 \log(\frac{2}{\delta})}{8 \beta_i^2 \alpha_i^2 }},
    \end{equation}

\begin{IEEEproof}
    Refer to Appendix~A.
\end{IEEEproof}
 \end{thm}
\begin{rem} \label{remark:rem1}
\textit{Theorem~\ref{thm:core_motivation} indicates that the generalization gap of the learned model on the unseen data mainly comes from two parts, including the complexity of hypothesis space, and the term $\sqrt{\sum _{i \in \boldsymbol{S}}\frac{ A^2 \log(\frac{2}{\delta})}{8 \beta_i^2 \alpha_i^2}}$, where $A$ represents the summation of concentration parameters of each user's $a_i$ and $\delta \in (0,1)$. 
To enhance generalization performance, the second term should be minimized. This can be achieved in the following ways:
\begin{enumerate}
    \item A skewed distribution of $a_i$ results in a smaller value for the second term compared to a uniform distribution of $a_i$ under the non-IID case;
    \item Involving more users and/or data in the training process.  
\end{enumerate}
}
\end{rem}
 
 Based on Remark~\ref{remark:rem1}, the size of the data can be increased by increasing the number of participating users. On the other hand, all users have distinct data distributions characterized by different values of $\alpha$. This makes it essential to use clustering algorithms to group users into separate clusters, which in turn helps to reduce the generalization gap.
In order to preserve data privacy in FL and save computational and transmission costs, the proposed framework aims to identify the cluster structure that best represents the entirety of the dataset, relying solely on the underlying distribution of each user.
\vspace{2.0mm}
\setlength{\textfloatsep}{2pt}
Denote $n_j$ by the number of appearances of $j$-th category belonging to a given local dataset, $\forall j = \{1, \cdots, C\}$. $p(n)$ represents the possibility of the equivalence class of all users that have the same label counts for the given data of one user. It can be derived from taking the integral of the product of the Multinomial likelihood and the Dirichlet prior. The expression can be obtained through manipulations as follows:
\begin{equation} \label{eq:pmf_md}
    p(n)=\frac{V !}{\prod_{j=1}^C n_{j} !} \frac{\Gamma(\alpha_0)}{\Gamma(\alpha_0+V)} \prod_{j=1}^C \frac{\Gamma\left(\alpha_j+n_{j}\right)}{\Gamma\left(\alpha_j\right)},
\end{equation}
where $V=\sum_{j=1}^C n_j$ denotes the total count of all categories occurrences for a given device, $\alpha_0 = \sum_{j=1}^C \alpha_j $ denotes the sum of the concentration parameters in Dirichlet distribution, and $\Gamma(\cdot)$ represents the gamma function, i.e., $\Gamma(n) = (n-1)!$.

The log-likelihood function of~(\ref{eq:pmf_md}) can be obtained by taking the logarithm of the last two terms, where the constant term $\frac{V !}{\prod_{j=1}^C n_{j} !}$ is ignored. Therefore it can be obtained through manipulations and expressed as follows:
\begin{equation}\label{eq:log_likelihood_MD}
\begin{aligned}
\mathcal{L}(\boldsymbol{\alpha})= &  \ln \Gamma(\alpha_0) -\ln \Gamma(\alpha_0+V)\\
& +\sum \nolimits_{j=1}^C\left(\ln \Gamma\left(\alpha_j+n_j\right)-\ln \Gamma\left(\alpha_j\right)\right).
\end{aligned}
\end{equation}
Then the gradient of $\mathcal{L}(\boldsymbol{\alpha})$ can be used in optimization techniques, i.e., BFGS algorithm, to determine the parameters for the clustering purpose, which is discussed in Section~\ref{sec:FL_methods}. Following this, using the clustering outcomes as a basis, the standard FL procedure is carried out within each cluster, utilizing the global models inherent to that particular cluster.

Within each cluster, the general supervised machine learning tasks are considered. The sample-wise loss function can be defined as $\ell_f\left ({{{\boldsymbol {w}},{\boldsymbol {x}_{i,k}},{y_{i,k}}} }\right)$ to quantify the prediction error between the data sample on learning model $\boldsymbol{w}$ and the ground truth label, where $\boldsymbol{x}_{i,k}$ and $y_{i,k}$ respectively denote the $k$-th input data of the $i$-th user and its corresponding ground truth label. The local loss function can be expressed as follows \cite{joint_communications_FL}:
\begin{equation} \label{eq:local_loss}
f_i(\boldsymbol{w}) = \frac {1}{\beta_i}  \sum \limits _{k=1 }^{\beta_{i}} {\ell_f\left ({{{\boldsymbol {w}},{ {\boldsymbol{x}}_{i,k}},{y_{i,k}}} }\right)}.
\end{equation}

Based on the local loss function, the global loss function $F({\boldsymbol {w}})$ can be expressed as follows:
\begin{equation} \label{eq:global_loss}
 F({\boldsymbol {w}}) =  \sum \limits _{i\in\boldsymbol {S} } \frac {\beta_i}{ \sum \limits _{i\in\boldsymbol {S} } \beta_i}   f_i(\boldsymbol{w}).
\end{equation}

\begin{Assume}\label{assumption:Lip}
\textit{
  The gradient $\nabla F\left(\boldsymbol{w}\right)$ is uniformly Lipschitz continuous with respect to $\boldsymbol{w}$, i.e.,
    \begin{equation}
     \left|\nabla F\left(\boldsymbol{w}^\mathrm{t}\right)-\nabla F\left(\boldsymbol{w}^{\mathrm{t}-1}\right)\right| \leq L\left|\boldsymbol{w}^\mathrm{t}-\boldsymbol{w}^{\mathrm{t}-1}\right| \text {, }
\end{equation}
where $\boldsymbol{w}^\mathrm{t}$ denotes the model parameters in the round $t$, $\left|\boldsymbol{w}^\mathrm{t}-\boldsymbol{w}^{\mathrm{t}-1}\right|$ is the norm of $\boldsymbol{w}^\mathrm{t}-\boldsymbol{w}^{\mathrm{t}-1}$, $L$ denotes the Lipschitz constant.
}
\end{Assume}

\begin{Assume}
    \textit{For the global loss function, the  $\mu$-PL inequality holds if there exists a constant  $\mu > 0$,the following inequality is satisfied:}

\begin{equation}
\frac{1}{2} \left|\nabla F(\boldsymbol{w})\right|^2 \geq \mu \left( F(\boldsymbol{w}) - F(\boldsymbol{w}^*) \right),
\end{equation}
\end{Assume}
Note the assumption outlined above is much less stringent than the traditional concept of strong convexity, and it is applicable to certain non-convex functions as well.
\begin{thm}\label{thm:convergence}
    \textit{Given the learning rate $\eta = \frac{1}{L}$, the expected reduction of global loss in round $t$ with an arbitrary set of users is bounded by}: 
    \begin{equation}
\begin{aligned}
&\mathbb{E}\left[F\left(\boldsymbol{w}^{\mathrm{t}}\right) - F\left(\boldsymbol{w}^*\right) \right] \\
\leq & \left(1 - \frac{\eta 2\mu}{\left(\sum_{i \in \boldsymbol{S}_{t-1}} \beta_i\right)^2} \sum_{i \in \boldsymbol{S}_{t-1}} \beta_i^2\right) \mathbb{E}\left[F\left(\boldsymbol{w}^{t-1}\right) - F\left(\boldsymbol{w}^*\right)\right] \\
& - \frac{\eta}{\left(\sum_{i \in \boldsymbol{S}_{t-1}} \beta_i\right)^2} \sum_{i \in \boldsymbol{S}_{t-1}} \beta_i^2 G^2,
\end{aligned}
\end{equation}
    where $\boldsymbol{S}_{t}$ denotes the set of selected users in round $t$.
\end{thm}
\begin{IEEEproof}
    Refer to Appendix B.
\end{IEEEproof}
Based on the definitions above, the global loss minimization problem can be formulated as follows:
\begin{align*} 
\label{equ:problem_formulation}
&\min \limits _{\mathcal {C}} F(\boldsymbol{w})\tag{P1}
\\ &~\,\rm {s.\;t.}\;
\mathcal {C}_{z} \cap \mathcal {C}_{z'} = \emptyset, \mathcal{C}_z, \mathcal{C}_{z'} \in \mathcal{C}, z\neq z',\tag{P1a}
\\ &\hphantom {~\,\rm {s.\;t.}\;}
{ \operatorname{card}\Bigl(\bigcup \nolimits_{i \in \mathcal{N}} c_{i, j}\Bigl) \leq C, \forall j \in\{1 \ldots C\},}\tag{P1b}
\\ &\hphantom {~\,\rm {s.\;t.}\;}
\end{align*}
where $\mathcal{C}$ denotes the set of cluster index sets, $\mathcal{C} = \{ \mathcal{C}_1, \cdots, \mathcal{C}_Z \}$, $c_{i, j}$ denotes the number of data points related to the 
$j$-th class in $i$-th user dataset, while $Z$ and $\mathcal{C}_z$ respectively denote the number of clusters and the set of users in $z$-th cluster $\mathcal{C}_z$.  (P1a) represents that each user can be only included in one cluster. (P1b) represents the training dataset of each user covers at most $C$ categories. 
It should be noted that the procedures within FL can be seamlessly carried out after the clustering process, and will not affect wireless transmission.

\subsection{Hybrid NOMA Transmission Model}
In the conventional OMA-based FL framework, each channel is utilized by one user, which limits the number of selected users in each communication round. However, multiple users can share the uplink channel simultaneously with the NOMA scheme, enabling a more efficient and faster FL process \cite{kaidi_AOI_FL}. 
 \begin{figure}[!htbp] 
    \centering
    \label{Fig:Hybrid}{%
       \includegraphics[width=0.9\linewidth]{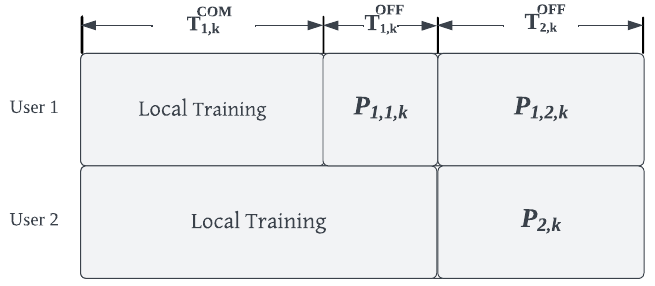}}
    \\
  \caption{Illustration of the considered hybrid NOMA system.}
  \label{fig:hybrid_NOMA} 
\end{figure}
In this paper, the hybrid NOMA scheme is considered \cite{hybrid_NOMA}. An illustration is shown in Fig.~\ref{fig:hybrid_NOMA}, where each sub-channel is occupied by two users. Since users have different computational capacities and data sizes, the user with less training time can transmit the local model first with the OMA scheme. The corresponding achievable data rate within the OMA scheme can be expressed as follows:
\begin{equation}
    R_{1,1,k} = B {\log_2}\left( {1 + {P_{1,1,k}{|h_{1,k}|^2}}}\right),
\end{equation}
where $B$ denotes the bandwidth of each sub-channel, $P_{1,1,k}$ denotes the transmit power of the user with less training time within the OMA scheme. Denote $|h_{1,k}|^2$ as the normalized channel condition, i.e., $|h_{1,k}|^2=|\hat{h}_{1,k}|^2 / \sigma^2$, where $|\hat{h}_{1,k}|^2$ denotes the complex channel
between the server and user via sub-channel $k$,
and $\sigma^2$ is the variance of the additive white Gaussian noise (AWGN). $h_{1,k} = L_{1} h_0$ denotes the channel of the first decoded user in sub-channel $k$. $L_{1} =\frac{\sqrt{\delta_1} \lambda \pi^{-1}}{4 d_1^{\alpha_{\text{PL}} / 2}}$ denotes large-scale fading \cite{NOMA_FL_cost}, where $\delta_1$, $\lambda$, $d_1$ and $\alpha_{\text{PL}}$ denote the antenna gain, signal wavelength, the distance between the server and first decoded user, and path loss exponent, respectively. $h_0$ denotes small-scale fading variable, i.e., $h_0 \sim \mathcal{CN}(0,1)$. Once the other user within the same sub-channel completes the local training, then the model parameters can be transmitted simultaneously with the NOMA scheme. Successive interference cancellation (SIC) is adopted for decoding, where the signal of the user with better channel quality is decoded first by treating the signal from another user as interference. Without loss of generality, it is assumed that $|h_{1,k}| \geq |h_{2,k}|$, and achievable data rate for the user with less training time in the NOMA scheme can be expressed as follows:
\begin{equation} 
R_{1,2,k} = B {\log_2}\left({1 + \frac {{P_{1,2,k}{|h_{1,k}|^2}}}{{ {P_{2,k}{|h_{2,k}|^2}} + 1}}}\right), 
\end{equation}
where $P_{1,2,k}$ and $P_{2,k}$ are the transmit power of the allocated users in $k$-th sub-channel transmitted within the NOMA scheme, and $h_{2,k}$ denotes the channel of the user with high training time in this case. 
In this work, the assumption is leveraged that the energy consumption of each device is mainly made up of two tasks, including the local training and the FL model transmission. Therefore, the overall energy consumption of the user with lower training time in sub-channel $k$ can be expressed as follows \cite{kaidi_AOI_FL,NOMA_FL_cost}:
\begin{equation} \label{eq:energy_consumption}
    E_{1,k} = \kappa \varsigma \vartheta_1 ^{2} \beta_1 + P_{1,1,k} T^\text{OFF}_{1,k} + P_{1,2,k} T^\text{OFF}_{2,k},
\end{equation}
where $\kappa$ is the energy consumption coefficient of CPU,  $\varsigma$ and $\vartheta_1$ respectively denote the number of CPU cycles required for computing per bit data and the frequency of the CPU clock on the device equipped on the user with lower training time, $T^\text{OFF}_{1,k}$ and $T^\text{OFF}_{2,k}$ respectively denote the offloading time of the first user transmitting part of the local model $D_1$ in OMA scheme and second user in NOMA scheme via sub-channel $k$. The corresponding expressions can be presented as $ \frac{D_1}{R_{1,1,k}}$ and $ \frac{D}{R_{2,k}}$, respectively. For the local training part, the computational energy consumption of two users in sub-channel $k$ can be presented as $E^{\text{COM}} = \kappa T_{1,k}^{\text{COM}} \vartheta_1 ^3 +\kappa T_{2,k}^{\text{COM}} \vartheta_2 ^3$. In this scenario, the overall energy consumption of the user with higher training time can be presented as $E_{2,k} = \kappa \varsigma \vartheta_2 ^{2} \beta_2 + P_{2,k} T^\text{OFF}_{2,k}$. Here, the signal is decoded last with the offloading data rate of $R_{2,k} =  B {\log_2}\left( {1 + {P_{2,k}{|h_{2,k}|^2}}}\right)$, which means that it is decoded without interference in the OMA scheme.

\section{Problem Formulation} \label{section:3}
In order to improve the performance of FL in the proposed framework, global loss minimization is chosen as the target. Following this, the problem of minimizing energy consumption will be detailed and formulated.

\subsection{Formulation of Energy Consumption Minimization}\label{subsec:formulation_subchannel}
The energy consumption minimization problem can be formulated as follows: 
\begin{align*}\label{eq:P2}
&\min_{\boldsymbol{s}, \mathbf{P}} \sum \nolimits_{k=1}^K \sum \nolimits_{i=1}^N s_{i, k}E_{i,k} ,\tag{P2}
\\ 
&\mathrm{s.t.}\quad s_{i, k} \in\{0,1\}, \quad \forall i \in \mathcal{N}, \forall k \in \mathcal{K}, \tag{P2a}
\\ &\hphantom {\,\,\,\rm {s.\;t.}\;}{\sum \nolimits_{i=1}^N s_{i, k} = 2,  \forall k \in \mathcal {K}},\tag{P2b}
\\ &\hphantom {\,\,\,\rm {s.\;t.}\;}{\sum \nolimits_{k=1}^K s_{i, k} = 1, \forall i \in \mathcal {N}},\tag{P2c}
\\ &\hphantom {\,\,\,\rm {s.\;t.}\;} 0 \leq P_{1,1,k} \leq P_{\mathrm{max1}}, 0 \leq P_{1,2,k}  \leq P_{\mathrm{max1}}, \tag{P2d}
\\ &\hphantom {\,\,\,\mathrm{s.t.}\;}0 \leq P_{2,k}\leq P_{\mathrm{max2}},\tag{P2e}
\\ &\hphantom {\,\,\,\mathrm{s.t.}\;}R_{2,k} T_{2,k}^{\text{OFF}} \geq D, \tag{P2f}
\\ &\hphantom {\,\,\,\mathrm{s.t.}\;} R_{1,1,k} T_d^{\text{OFF}} + R_{1,2,k} T_{2,k}^{\text{OFF}} \geq D, \tag{P2g} 
&\phantom{\mathrm{s.t.}}
\end{align*}
where $\boldsymbol{s}$ denotes the collection of sub-channel allocation indicators $s_{i,k}$. (P2a) is the constraint of sub-channel allocation indicators. (P2b) and (P2c) represent that two users are allocated to each sub-channel, and each user can only be assigned to one sub-channel, respectively. 
$P_{\text{max1}}$ and $P_{\text{max2}}$ respectively denote the maximum powers of two users. Constraints (P2d) and (P2e) indicate the requirements of the transmit power. (P2f) and (P2g) represent the requirements of the data transmission. 

The problem of energy consumption minimization is further divided into two sub-problems, including the optimization of sub-channel and power allocation.
Note that the sub-problems maintain direct relevance to the main problem (\ref{equ:problem_formulation}) through its critical role in determining the effective utility of the function $F(\boldsymbol{w})$. This is because the optimal resource allocation (\ref{eq:P2}) is achieved through minimizing the function in (\ref{eq:P3}) and (\ref{equ:P4}), which promotes more efficient user engagement, thereby benefiting the minimization of $F(\boldsymbol{w})$.
Therefore, the optimal solutions of (\ref{eq:P3}) and (\ref{equ:P4}) are not just a pathway to optimize energy consumption but also serve as a pivotal step towards achieving the global objective of minimizing the global loss in (\ref{equ:problem_formulation}). The sub-channel allocation problem can be formulated as follows: 
\begin{align*}\label{eq:P3}
&\min_{\boldsymbol{s}} \sum \nolimits_{k=1}^K \sum \nolimits_{i=1}^N s_{i, k}E_{i,k} ,\tag{P3}
\\ 
&\mathrm{s.t.}\quad (\text{P2a}),(\text{P2b}),(\text{P2c}),\tag{P3a}
&\phantom{\mathrm{s.t.}}
\end{align*}
It is vital to address (\ref{eq:P3}) before solving (\ref{equ:P4}), as the sub-channel allocation determined in (\ref{eq:P3}) directly influences the feasible solution of the power allocation in (\ref{equ:P4}), which means that both methods are solved iteratively. Following the sub-channel allocation, the power allocation is utilized to optimize energy consumption. This optimization problem is solved using the KKT conditions, and the problem can be formulated as follows:

\begin{align*} \label{equ:P4}
&\min _{\mathbf{P}} P_{1,1,k} T_{1,k}^{\text{OFF}} + P_{1,2,k} T_{2,k}^{\text{OFF}}  + P_{2,k} T_{2,k}^{\text{OFF}} + E^{\text{COM}},\tag{P4}
\\ 
&\mathrm{s.t.}\quad (\text{P2d})-(\text{P2g}),\tag{P4a}
\\ &\hphantom {\,\,\,\mathrm{s.t.}\;} T_{1,k}^{\text{COM}} + T_{1,k}^{\text{OFF}} + T_{2,k}^{\text{OFF}} \leq T_{\mathrm{max}}, T_{2,k}^{\text{COM}} + T_{2,k}^{\text{OFF}} \leq T_{\mathrm{max}},\tag{P4b}
\end{align*}
(P4b) indicates that the time constraints for both users.

\section{Description of Methodologies in CFL Framework}\label{sec:FL_methods}
\subsection{BFGS Algorithm in Log-Likelihood} 

 To estimate the concentration parameters, the log-likelihood function in~(\ref{eq:log_likelihood_MD}) is aimed to be maximized. For this purpose, the BFGS algorithm is employed. This algorithm is a Quasi-Newton method designed for solving unconstrained nonlinear optimization problems and is well-known for its properties, including no need for computation of Hessian matrix and its overall robustness. Denote the matrix of data appearances of all devices by $\boldsymbol{X}$. The core rationale can be described in the following: an initial guess for the concentration parameters is chosen, along with an initial approximation for the inverse Hessian matrix $\boldsymbol{H}^{(0)}$, which is commonly set to the identity matrix. In each iteration $t$, the gradient $\nabla \mathcal{L}(\boldsymbol{\alpha}^{(t)})$ is computed for calculating the search direction $\boldsymbol{p}^{(t)}$ using the current inverse Hessian approximation $\boldsymbol{H}^{(t)}$. 
 A line search along the search direction $\boldsymbol{p}^{(t)}$ is then used to find the optimal step size $\boldsymbol{s}^{(t)}$, leading to update the concentration parameter $
\boldsymbol{\alpha}^{(t+1)} = \boldsymbol{\alpha}^{(t)} + \boldsymbol{s}^{(t)}\boldsymbol{p}^{(t)}$. 
The approximation $\boldsymbol{H}^{(t+1)}$ is subsequently updated through $\boldsymbol{s}^{(t)} = \boldsymbol{\alpha}^{(t+1)} - \boldsymbol{\alpha}^{(t)}$ and the difference in the gradients $\boldsymbol{y}^{(t)} = \nabla \mathcal{L}(\boldsymbol{\alpha}^{(t+1)}) - \nabla \mathcal{L}(\boldsymbol{\alpha}^{(t)})$. The detailed expression of $\boldsymbol{H}^{(t+1)}$ can be found in \textbf{Algorithm}~\ref{alg:L_BFGS}. This iterative process continues until the convergence criteria is met or equivalently. 
The gradient of the $\mathcal{L}(\boldsymbol{\alpha})$ with respect to  $\alpha_j$ can be obtained through (\ref{eq:log_likelihood_MD}), and it can be expressed as follows:

\begin{equation}
\begin{aligned}
\frac{\partial \ln \mathcal{L}(\boldsymbol{\alpha})}{\partial \alpha_j} = &\psi(\alpha_0) -\psi(\alpha_0+V) + \psi(\alpha_j+n_j) - \psi(\alpha_j),
    \end{aligned}
\end{equation}
where $\psi(\cdot)$ is the digamma function, which is the derivative of the logarithm of the gamma function.
The main steps of the BFGS algorithm are given
in \textbf{Algorithm}~\ref{alg:L_BFGS}. 
 \begin{algorithm}
\caption{BFGS algorithm for Log-Likelihood Maximization}
\label{alg:L_BFGS} 
    \textbf{Input:} Data matrix $\boldsymbol{X}$, initial guess \( \boldsymbol{\alpha}^{(0)} \), tolerance \( \epsilon \)\\
    \textbf{Output:} Estimated concentration parameter \( \boldsymbol{\alpha}^* \)\\
     Set \( \boldsymbol{\alpha}^{(t)} = \boldsymbol{\alpha}^{(0)} \), \( t = 0 \), and initialize the inverse Hessian approximation \( \boldsymbol{H}^{(0)} \)\\
    \While{not converged}{
         Compute the gradient \( \nabla \mathcal{L}(\boldsymbol{\alpha}^{(t)}) \)\\
         Calculate the search direction \( \boldsymbol{p}^{(t)} = -\boldsymbol{H}^{(t)}\nabla \mathcal{L}(\boldsymbol{\alpha}^{(t)}) \)\\
         Perform line search to find the optimal \( \boldsymbol{s}^{(t)} \)\\
         Update the concentration parameter: \( \boldsymbol{\alpha}^{(t+1)} = \boldsymbol{\alpha}^{(t)} + s^{(t)}\boldsymbol{p}^{(t)} \)\\
         Compute \( \boldsymbol{s}^{(t)} = \boldsymbol{\alpha}^{(t+1)} - \boldsymbol{\alpha}^{(t)} \)\\
         Compute \( \boldsymbol{y}^{(t)} = \nabla \mathcal{L}(\boldsymbol{\alpha}^{(t+1)}) - \nabla \mathcal{L}(\boldsymbol{\alpha}^{(t)}) \)\\
         Update the inverse Hessian approximation \( \boldsymbol{H}^{(t+1)} \):
        \begin{equation}
        \begin{aligned}
            \boldsymbol{H}^{(t+1)} = & \left(\boldsymbol{I} - \frac{\boldsymbol{s}^{(t)}{\boldsymbol{y}^{(t)T}}}{\boldsymbol{y}^{(t)T}\boldsymbol{s}^{(t)}}\right)\boldsymbol{H}^{(t)}\left(\boldsymbol{I} -\frac{\boldsymbol{y}^{(t)}{\boldsymbol{s}^{(t)T}}}{\boldsymbol{y}^{(t)T}\boldsymbol{s}^{(t)}}\right) \\
            & + \frac{\boldsymbol{s}^{(t)}{\boldsymbol{s}^{(t)T}}}{\boldsymbol{y}^{(t)T}\boldsymbol{s}^{(t)}}.
        \end{aligned}
        \end{equation}\\
         Check for convergence: If \( ||\nabla \mathcal{L}(\boldsymbol{\alpha}^{(t+1)})|| < \epsilon \) or \( ||\boldsymbol{\alpha}^{(t+1)} - \boldsymbol{\alpha}^{(t)}|| < \epsilon \), then converged\\
         Set \( t = t + 1 \)\\
    }
     \( \boldsymbol{\alpha}^* = \boldsymbol{\alpha}^{(t)} \)\\
\end{algorithm}
\setlength{\textfloatsep}{0pt}
The optimization problem can be transformed to minimize the negative of the log-likelihood function, i.e., $\boldsymbol{\alpha}^* = \underset{\boldsymbol{\alpha}}{\operatorname{argmax}} \mathcal{L}(\boldsymbol{\alpha}) = \underset{\boldsymbol{\alpha}}{\operatorname{argmin}} -\mathcal{L}(\boldsymbol{\alpha})$, subject to $\alpha_j > 0$.
The estimated concentration parameter $\boldsymbol{\alpha}^*$ is obtained when the algorithm converges. Convergence can be determined based on various criteria, such as a maximum number of iterations, the gradient magnitude falling below a certain threshold, or the difference in function values between consecutive iterations being smaller than a predefined tolerance as shown in step 12 of \textbf{Algorithm}~\ref{alg:L_BFGS}.

\subsubsection{Complexity Analysis} 
Operations involved in the BFGS algorithm for the proposed framework are investigated for the complexity analysis. The overall complexity is mainly based on the steps of gradient computation, line search, and the inverse Hessian-vector product.
Evaluating the gradient requires $O(N_{\text{bfgs}})$ operations, where $N_{\text{bfgs}}$ is the number of parameters, which is the size of $\boldsymbol{X}$. The line search procedure requires multiple evaluations of the function with its gradient, resulting in multiple $O(N_{\text{bfgs}})$ computations. Assume the average number of evaluations in the line search is $U$, and if each evaluation takes $K$ iterations to converge, the total cost for this step becomes $O(UN_{\text{bfgs}}K)$.
The core of the BFGS algorithm is the calculation of the approximation to the inverse Hessian-vector product. This step considers the vector-matrix multiplication and outer product, which requires $O(N_{\text{bfgs}}^2)$ operations.
Assuming the algorithm converges after $T$ iterations, the total computational complexity of the proposed BFGS algorithm over all iterations can be represented as
$[O(T(UN_{\text{bfgs}}K + N_{\text{bfgs}}^2))]$, which can be approximated to $[O(T N_{ \text{bfgs}}^2)]$.

\subsection{Spectral Clustering} \label{subsection:Spectral}
Inspired by image segmentation, a clustering method based on spectral clustering with the relaxed RatioCut is proposed.
Spectral clustering is a well-known unsupervised learning algorithm and is extensively employed across various applications as it allows flexibility on the input data compared to traditional clustering algorithms, i.e., K-means\cite{spectral}. In this subsection, a brief description of the proposed spectral clustering is provided. For a more in-depth understanding of spectral clustering, the concept can refer to \cite{spectral_tutorial}.
First consider $\mathcal{G} = (\mathcal{V}, \mathcal{E},\mathbf{W})$ as the graph that contains vertices and edges constructed on the corresponding concentration parameters of devices aforementioned, where the vertex set denoted by $\mathcal{V}$, the set of all edges connecting the vertex denoted by $\mathcal{E}$, and the weight matrix calculated by weight function denoted by $\mathbf{W}$.  
The degree matrix $\mathbf{D}$ is defined as the diagonal matrix with the degrees $d_1, \cdots, d_n$ on its diagonal with corresponding weights $w_{ij}$ between two vertices $v_i$ and $v_j$, where the degree of a vertex is denoted by $d_i = \sum_{j=1} ^ n w_{ij}$.
Denote $\mathbf{L}$ by the $n\times n$ unnormalized graph Laplacian matrices where $n$ represents the size of data points. It can be obtained through $\mathbf{L} = \mathbf{D} - \mathbf{W}$, where $\mathbf{L}$ remains positive semi-definite.  Denote $\mathbf{A}=(\boldsymbol{\alpha}_1,\cdots,\boldsymbol{\alpha}_Z) \in \mathbb{R}^{n\times Z}$, where $\boldsymbol{\alpha}_z$ is the $z$-th column chosen from the eigenvectors of $\mathbf{L}$. 
The optimization objective of spectral clustering with relaxed RatioCut can be expressed as follows \cite{K_means_bound}:
\begin{equation}\label{eq:objective_spetral}
\begin{aligned}
\min_{\mathbf{A}} \quad & \mathcal{L}_s(\mathbf{A}) = C_s\sum_{z=1}^Z \sum_{i, j=1, i \neq j}^n \mathbf{W}_{i, j}\left(\alpha_{z,i}-\alpha_{z,j}\right)^2 \\
\text{s.t.} \quad & \mathbf{A}^T \mathbf{A}=\mathbf{I},
\end{aligned}
\end{equation}
where $C_s$ denotes $\frac{1}{2 n(n-1)}$, $\alpha_{z,i}$ is the $i$-th element of the corresponding the $z$-th eigenvector, and $\mathbf{I}$ is the identity matrix. It aims to cluster the data in such a way that the total edge weight connecting different clusters is minimized. 
The core concept of spectral clustering with relaxed RatioCut can be described as follows: A similarity matrix $\mathbf{W}$ is constructed by computing the pairwise similarities between data points, typically using a Gaussian kernel function \cite{K_means_bound}. From this matrix, $\mathbf{L}$ can be subsequently derived. The first $Z$ eigenvectors corresponding to the smallest non-zero eigenvalues of $\mathbf{L}$ are computed to form the columns of matrix  $\mathbf{A}$. Then the objective is to find an $\mathbf{A}$ that minimizes the function, reducing the total edge weight between different clusters, which can be solved by Rayleigh–Ritz method \cite{spectral_tutorial}. Then the rows of $\mathbf{A}$ are normalized to have unit length, forming a new representation in a low-dimensional space. These rows of $\mathbf{A}$ are subsequently treated as features in $\mathbb{R}^Z$ and gathered together using a clustering algorithm, i.e., K-means. The cluster assignments in the transformed space are then used to assign the original data points to the respective clusters. 
\subsubsection{Complexity Analysis}\label{Complexity:spectral}
Following the similar analysis provided in the BFGS algorithm, the overall complexity of procedures spectral clustering with relaxed RatioCut for the considered framework largely comes from eigendecomposition, which can be approximated to $O(n^3)$. The method like QR decomposition has $O(n^2)$ in general, which can be applied to further reduce the complexity. The servers usually possess superior computational capabilities, including the ability to perform parallel computations. This allows the proposed clustering solution to decrease computational overhead, considering a notable advantage of the proposed framework.

\subsubsection{Selecting Number of Clusters}\label{subsection:select_clusters}
The performance of the spectral clustering algorithm is influenced by the number of clusters. However, it is well known that determining the optimal number of clusters is NP-hard. This holds not just for spectral clustering but for clustering algorithms in general. Considering an extreme case under the proposed framework, the selected number of clusters would intuitively equal to the number of labels if users have data of only one label as shown in Fig.~\ref{fig1:user_distribution}(a). However, such a selection method may not always yield the optimal choice as the quantity of data could also influence the choice of the number of clusters. In order to gain more insights, the excess risk bounds of spectral clustering on relaxed Ratiocut $\mathcal{L}_s(\mathbf{A}^*)-\mathcal{L}_{sd}(A^*)$ can be analyzed. For any $\delta>0$, with probability at least $1-\delta$\cite{on_learning_operator}:
\begin{equation}\label{eq:spectal_generalization}
    \mathcal{L}_s(\mathbf{A}^*)-\mathcal{L}_{sd}(A^*) = 2Z \frac{ \sqrt{2} \kappa_s \sqrt{\log \frac{2}{\delta}}}{\sqrt{n}},
\end{equation}
where $\mathcal{L}_s(\mathbf{A}^*) - \mathcal{L}_{sd}(A^*)$ denotes the gap between the empirical solution on the optimization objective in~(\ref{eq:objective_spetral}) and its optimal solution in continuous domain related to the size of the data points $n$. $\mathbf{A}^* = (\boldsymbol{\alpha_1^*},\cdots,\boldsymbol{\alpha}_Z^*)$ and $A^*= (\alpha^*_1,\cdots,\alpha^*_Z)$  represent the optimal solutions of the discrete and continuous version of~(\ref{eq:objective_spetral}), each element in $\mathbf{A}^*$ and $A^*$ corresponds to the eigenvectors of Laplacian $\mathbf{L}$ and operator $L_Z$, where $L_Z$ is the kernel function defined for the continuous version of~(\ref{eq:objective_spetral}) that is associated with the reproducing kernel Hilbert space\footnote{It is important to note that $\mathbf{A}^*$ and $A^*$ are in different spaces, where $\mathbf{A}^*$ is in the finite-dimensional space, and the $A^*$ is in infinite-dimensional space. The concept can refer to \cite{on_learning_operator} for more insights.}.
$\kappa_s$ denotes the upper bound of the kernel function $L_Z$. 

Based on the (\ref{eq:spectal_generalization}), it is evident that the generalization performance of spectral clustering on Relaxed Ratiocut is decreased with an increased number of clusters $Z$, but improved proportionally to the square root of the size of the data points. It indicates the importance of carefully selecting the number of clusters and also guides the design of future works, i.e., how the proposed method generalizes when the new user joins.
One heuristic method to obtain $Z$ is to evaluate the eigengap for each eigenvalue. Denote $z$-th eigenvalue by $\uplambda_z$, where the second eigenvalue, i.e., $z = 2$. The optimal $Z$ is chosen where the eigengap is maximized \cite{select_k_spectal}: $\underset{Z}{\operatorname{argmax}}\left(\uplambda_z-\uplambda_{z+1}\right)$. The eigengap indicates a transition from these tightly packed clusters to more dispersed clusters, suggesting an optimal stopping point for the number of clusters. Consequently, if a distinct eigengap is discernible, then the number of this cluster corresponding to the gap will be the optimal $Z$. 
However, there may be instances where multiple gaps or none at all will be observed. In these scenarios, a brute-force search approach is employed to determine the most suitable $Z$. The corresponding simulation is illustrated in Fig.~\ref{fig:clusters}.

\section{Iterative Optimization of Sub-Channel and
Power Allocation}\label{sec:wireless_methods}

\subsection{Matching Theory-based Sub-Channel Allocation}
In this subsection, the solution of sub-channel assignment is proposed.
The sets of users within cluster $\mathcal{C}_z$ and sub-channels $\mathcal{K}$ are considered as two disjoint sets of players, i.e., $\mathcal{C}_z \cap \mathcal{K}=\emptyset$. By allocating any user $m\in \mathcal{C}_z$ to a sub-channel $k \in \mathcal{K}$, the sub-channel allocation problem can be described as a two-to-one matching, in which the definition can be defined below.

\begin{defn}
\textit{
A two-to-one matching $\mu$ is a mapping function from the set $\mathcal{C}_z \cup \mathcal{K}$ into the set of all elements of $\mathcal{C}_z \cup \mathcal{K}$ such that:}
    
1) $\mu(m) \in \mathcal{K}, \forall m \in \mathcal{C}_z, \mu(k) \subseteq \mathcal{C}_z, \forall k \in \mathcal{K}$;

2) $|\mu(m)| = 1, \forall m \in \mathcal{C}_z,|\mu(k)| = 2, \forall k \in \mathcal{K}$;

3) $m\in \mu(k) \Leftrightarrow \mu(m)=k$.

\end{defn}

The above definition states that: 
1) any user $m$ is matched with one of the sub-channels and any sub-channel $k$ is matched with a subset of users.  2) each user is matched with one sub-channel, and each sub-channel is matched with two users. 3) user $m$ and sub-channel $k$ are matched with each other. 

\begin{algorithm}
\DontPrintSemicolon
\caption{Sub-Channel Allocation Algorithm}
\label{algorithm:matching}
Randomly initialize matching $\mu$. \\
\While{no further swap-blocking pair}{
Any user $m$ searches another user $j$, where $\mu(j) \neq \mu(m)$. \\
\If{$(j, m)$ is a swap-blocking pair}{

\quad 1) Matching $\mu_m^j$ is approved. \\
\quad 2) Sub-channels swap between user $m$ and $j$.\\
\quad 3) Set $\mu=\mu_m^j$. 
}
\Else{
Keeps current the matching $\mu$.
     }
}
\end{algorithm}
\setlength{\textfloatsep}{0pt}
The model assumes that users cannot be left unmatched. This aligns with the notion of two-sided exchange matchings as described in prior research\cite{FL_fading_channels, kaidi_AOI_FL}.
In a two-sided exchange matching model, a swap operation involving any user between two different sub-channels is essentially the act of two users exchanging their assigned sub-channels. A swap matching $\mu_m^j$ implies that user $m$ switches to user $j$ 's sub-channel, and user $j$ is assigned to user $m$ 's sub-channel, while other users remain the same.

\begin{defn}
    A swap matching $\mu_m^j=$ $\left\{\mu \backslash\left\{(k, j),\left(k^{\prime}, m\right)\right\} \cup\left\{(k, m),\left(k^{\prime}, j\right)\right\}\right\}$, where $j \in \mu(k)$, $m\in \mu\left(k^{\prime}\right), j \in \mu_m^j\left(k^{\prime}\right)$, and $m\in \mu_m^j(k)$.
\end{defn} 
It is worth noting that the externalities exist, indicating that any swap operation is required to be approved by the players directly involved in the swap. For example, consider a swap operation of users $j$ and $m$, where $k=\mu(j), k^{\prime}=\mu(m)$, it results in a transformation of the matching from $\mu$ to $\mu_m^j$, where $k=\mu_m^j(m), k^{\prime}=\mu_m^j(j)$. Throughout this swap procedure, the energy consumption of any player will not be increased, and at least one of these players can reduce its energy consumption. Following a series of such swap operations, a two-sided exchange-stable matching is obtained in the considered problem.
\begin{defn}
\textit{$\mu$ satisfies two-sided exchange stability (2ES) if, and only if, there is no swap-blocking pair of users.}
\end{defn}
The matching-based sub-channel allocation algorithm aims to determine the $2\text{ES}$ structure within the formulated matching model can be proposed based on Definition 3. The details of the approach are described in \textbf{Algorithm}~\ref{algorithm:matching}.

\subsubsection{Complexity Analysis}
Each user must check all others who are allocated to different sub-channels in the worst case. Consequently, $N-2$ users are needed to execute the swap operations. When there are $N$ users, the calculations could at the most, be conducted $N(N-2)$ times within a full cycle. Given a specified number of cycles, denoted as $T_m$, the highest computational complexity that the \textbf{Algorithm}~\ref{algorithm:matching} can be approximated to $O(T_m N^2)$.

\subsection{Closed-Form Solution by KKT Conditions}
In this subsection, the formulated power allocation problem in~(\ref{equ:P4}) is addressed using the KKT conditions, for which the closed-form solutions are derived.
As proved in \cite{kaidi_AOI_FL}, the minimum energy consumption is only achieved when the computational and offloading time of each user equals the maximum tolerance time, i.e., $T_{m,k}^{\text{COM}}+T_{m,k}^{\text{OFF}} = T_{\text{max}}$, where $m \in \{1,2\}$.
It indicates that the transmit power $P_{2,k}$ of the user at the second decoded stage denoted by $P^*_{2,k}$, can be expressed as $(2^{\frac{ B^{-1} D }{ T_{2,k}^{\text{OFF}}} }-1)/\left|h_{2,k}\right|^2$, and constraint (P3c) and (P3d) are both equal to $D$. Then the following proposition can be further presented: 
\begin{prop}\label{prop:convex}
    Based on assumption $|h_{1,k}| \ge |h_{2,k}|$, the problem can be proved to be convex.
\end{prop}
\begin{IEEEproof}
For the formulated problem to be convex, the following sufficient conditions must be met:
1) minimization of a convex function, or maximization of a concave function; 2) each constraint describes a convex set. It can be seen that
(P3a) and (P3b) are linear. (P3c) exhibits convexity because its Hessian matrix is positive definite, which can be expressed as follows: 
\begin{equation}
\begin{bmatrix}
\frac{B T_d^{\text{{OFF}}} \left|h_{1,k}\right|^4}{(P_{1,k} \left|h_{1,k}\right|^2 + 1)^2} & 0 \\
0 & \frac{B T_{2,k}^{\text{{OFF}}} \left|h_{1,k}\right|^4}{(P_{2,k} \left|h_{2,k}\right|^2 + 1)^2 \left(\frac{P_{1,2,k} \left|h_{1,k}\right|^2}{P_{2,k} \left|h_{2,k}\right|^2 +1} + 1\right)^2}
\end{bmatrix}.
\end{equation}
Given that both the denominator and numerator of the elements on the diagonal of the Hessian matrix are non-zero, it states that the matrix is positive definite. The proof is completed.
\end{IEEEproof}
Due to the fact that the convexity and Slater's condition are satisfied, the solution can be obtained by using KKT conditions.
The Lagrangian function of the problem can be obtained as follows: 

\begin{equation}
\begin{aligned}
&\mathcal{L}_\text{KKT} = P_{1,1,k} T_d^{\text{OFF}} + P_{1,2,k} T_{2,k}^{\text{OFF}} + P_{2,k} T_{2,k}^{\text{OFF}} + E^{\text{COM}}\\
&+ \lambda_1(P_{1,1,k}-P_{max1}) + \lambda_2(-P_{1,1,k})+ \lambda_3(P_{1,2,k}-P_{max1})\\
& + \lambda_4(-P_{1,2,k})+ \mu_1\biggl(P_{2,k}-\frac{P^*_{2,k}-1}{\left|h_{2,k}\right|^2}\biggl)\\
& + \mu_2\Bigl(D - R_{1,1,k} T_d^{\text{OFF}} - R_{1,2,k} T_{2,k}^{\text{OFF}} \Bigl),
\end{aligned}
\end{equation}
where $\lambda_i$ and $\mu_i$ denote the Lagrangian multipliers for the inequality and equality constraints in~(\ref{equ:P4}), respectively. 
Moreover, the following conditions hold: 
\begin{equation}
\left\{\begin{aligned}
&P_{1,1,k} - P_{\text{max1}}\leq 0, P_{1,1,k} \geq 0,  & \text{(18a)}\\
&P_{1,2,k} - P_{\text{max1}} \leq 0, P_{1,2,k} \geq 0, P_{2,k} - P_{\text{max2}} \leq 0,  & \text{(18b)}\\
&\biggl(\frac{P^*_{2,k}-1}{\left|h_{2,k}\right|^2} - P_{2,k}\biggl)= 0, & \text{(18c)}\\
&D - R_{1,1,k} T_d^{\text{OFF}} - R_{1,2,k} T_{2,k}^{\text{OFF}} = 0, & \text{(18d)}\\
&\lambda_1 (P_{1,1,k} - P_{\text{max1}}) = 0, \lambda_2 (-P_{1,1,k}) = 0, & \text{(18e)}\\
&\lambda_3 (P_{1,2,k} - P_{\text{max1}}) = 0, \lambda_4 (-P_{1,2,k}) = 0, & \text{(18f)}\\
&\mu_1 \biggl(\frac{P^*_{2,k}-1}{\left|h_{2,k}\right|^2} - P_{2,k}\biggl) = 0, & \text{(18g)}\\
&\mu_2 (D - R_{1,1,k} T_d^{\text{OFF}} - R_{1,2,k} T_{2,k}^{\text{OFF}}) = 0, & \text{(18h)}\\
&\lambda_i \geq 0,  \forall i \in\{1,2,3,4\}, \quad \mu_i \geq 0,  \forall i \in\{1,2\}. & \text{(18i)}
    \end{aligned}\right.
\end{equation}
The gradient of each optimal value of transmit power can be expressed as follows: 
\begin{equation}
    \begin{aligned}
&\frac{\partial \mathcal{L}}{\partial P_{1,1,k}} = -\frac{B T_{d}^{\text{OFF}}  \left|h_{1,k}\right|^2 \mu_2}{P_{1,1,k} \left|h_{1,k}\right|^2 + 1} + T_{d}^{\text{OFF}} + \lambda_1 - \lambda_2
,\\
&\frac{\partial \mathcal{L}}{\partial P_{1,2,k}} = -\frac{B  T_{2,k}^{\text{OFF}} \left|h_{1,k}\right|^2  \mu_2}{(P_{2,k} \left|h_{2,k}\right|^2 + 1) (\frac{P_{1,2,k} \left|h_{1,k}\right|^2}{P_{2,k} \left|h_{2,k}\right|^2 + 1} + 1)} \\
& \quad\quad\quad\quad\quad\quad  + T_{2,k}^{\text{OFF}} + \lambda_3 - \lambda_4,
    \end{aligned}
\end{equation}

Through the manipulations, the optimal values of $P_{1,1,k}$ and $P_{1,2,k}$ can be expressed through the partial derivatives above, while the expression of the latter is expressed by~(\ref{eq:close_form_p12k}), as shown at the top of the next page:

\begin{equation}\label{eq:close_form_p11k}
P_{1,1,k} = \frac{B T_{d}^{\text{OFF}}  h_{1,k}^2 \mu_2 - T_{d}^{\text{OFF}} - \lambda_1 + \lambda_2}{h_{1,k}^2 (T_{d}^{\text{OFF}} + \lambda_1 - \lambda_2)},
\end{equation}
\begin{figure*}
\begin{equation}\label{eq:close_form_p12k}
P_{1,2,k} = \frac{B  T_{2,k}^{\text{OFF}}  \left|h_{1,k}\right|^2  \mu_2 - P_{2,k}  T_{2,k}^{\text{OFF}}  \left|h_{2,k}\right|^2 - P_{2,k}  \left|h_{2,k}\right|^2  \lambda_3 + P_{2,k}  \left|h_{2,k}\right|^2  \lambda_4 - T_{2,k}^{\text{OFF}} - \lambda_3 + \lambda_4}{\left|h_{1,k}\right|^2  (T_{2,k}^{\text{OFF}} + \lambda_3 - \lambda_4)},
\end{equation}
\hrulefill
\end{figure*}
Based on the above conditions and equations, the following proposition is obtained for the closed-form solutions of~(\ref{equ:P4}):

\begin{prop}
    \textit{Denote $A_1 = P_{1,1,k} |h_{1,k}|^2 + 1$, $B_1 = P_{2,k} \left|h_{2,k}\right|^2 + 1$, and $B_2 = P_{max1} |h_{1,k}|^2 + 1$. The closed-form solutions based on different Lagrangian multipliers are provided, which can be expressed as follows}:
\begin{enumerate}
    \item 
    \textit{$\lambda_1,\lambda_2 = 0, \mu_2 > 0$}:
        \begin{equation}
        \begin{aligned}
        &P_{1,1,k} = \frac{-1 + \exp \Bigl(- \log \Bigl(\frac{P_{1,2,k} \left|h_{1,k}\right|^2 + B_1}{B_1}\Bigl) - D B^{-1} \Bigl)}{\left|h_{1,k}\right|^2},\\
        &P_{1,2,k} = \Bigl(-1 + \exp\Bigl(-\frac{B \log\left(A_1\right) - D}{B}\Bigl)\Bigl)\Bigl(\frac{B_1}{|h_{1,k}|^2}\Bigl),\\
        \end{aligned}
        \end{equation}
    \item 
    \textit{$\lambda_1 = 0, \lambda_2 > 0, \mu_2 > 0$}:
        \begin{equation}
        \begin{aligned}        
        &P_{1,1,k} = 0,\\
        &P_{1,2,k} = \frac{B_1 (\exp(\frac{D}{B}) - 1)}{|h_{1,k}|^2},
        \end{aligned}
        \end{equation}
    \item 
    \textit{$\lambda_1 > 0, \lambda_2 = 0, \mu_2 > 0$}:
        \begin{equation}
        \begin{aligned}        
        &P_{1,1,k} = P_{max1},\\
        &P_{1,2,k} = \Bigl(-1 + \exp\Bigl(- \log\left(B_2\right) - D B^{-1}\Bigl)\Bigl)\Bigl(\frac{B_1}{|h_{1,k}|^2}\Bigl),
        \end{aligned}
        \end{equation}
        
\end{enumerate}
\end{prop}
\begin{IEEEproof}
    Given space limitations, the detailed derivation of these solutions is omitted, instead, a brief overview is provided. Based on~(\ref{eq:close_form_p11k}) and~(\ref{eq:close_form_p12k}), the possible closed-form solution of problem~(\ref{equ:P4}) can be categorized into 8 cases, each determined by the different Lagrangian multipliers. These categories can be further validated based on the valid combinations, i.e., it can be noted that the case when $\mu_2 = 0$ is not valid. The proof is completed.
\end{IEEEproof}

\section{Simulation Results} \label{section:simulation}
In this section, the simulation results are presented to demonstrate the performance of the proposed solutions. For fair comparisons, each simulation related to FL is optimized over learning rates in two orders of magnitude, i.e., $\eta \in \{10^{-3},5\times10^{-3},\cdots,10^{-1},5\times10^{-1}\}$ in MNIST, and $\eta \in \{10^{-4},5\times10^{-4},\cdots,10^{-2},5\times10^{-2}\}$ in CIFAR-10 dataset, with each learning rate averaged over 3 runs. For all conducted simulations, $\kappa = 10^{-28}$, $\varsigma = 10^7$, $\vartheta \in [1.8 \times 10^9, 2.2 \times 10^9]$, $D = 1.1$ Mbit, $\alpha_{\text{PL}} = 2$, radius of the disc = $600$ m, and $B = 1$ MHz \cite{kaidi_AOI_FL, NOMA_FL_cost}.
\begin{figure}[!htbp]
    \centering
    \subfloat[Test accuracy on MNIST.\label{Fig:output_mnist}]{%
        \includegraphics[width=0.9\linewidth]{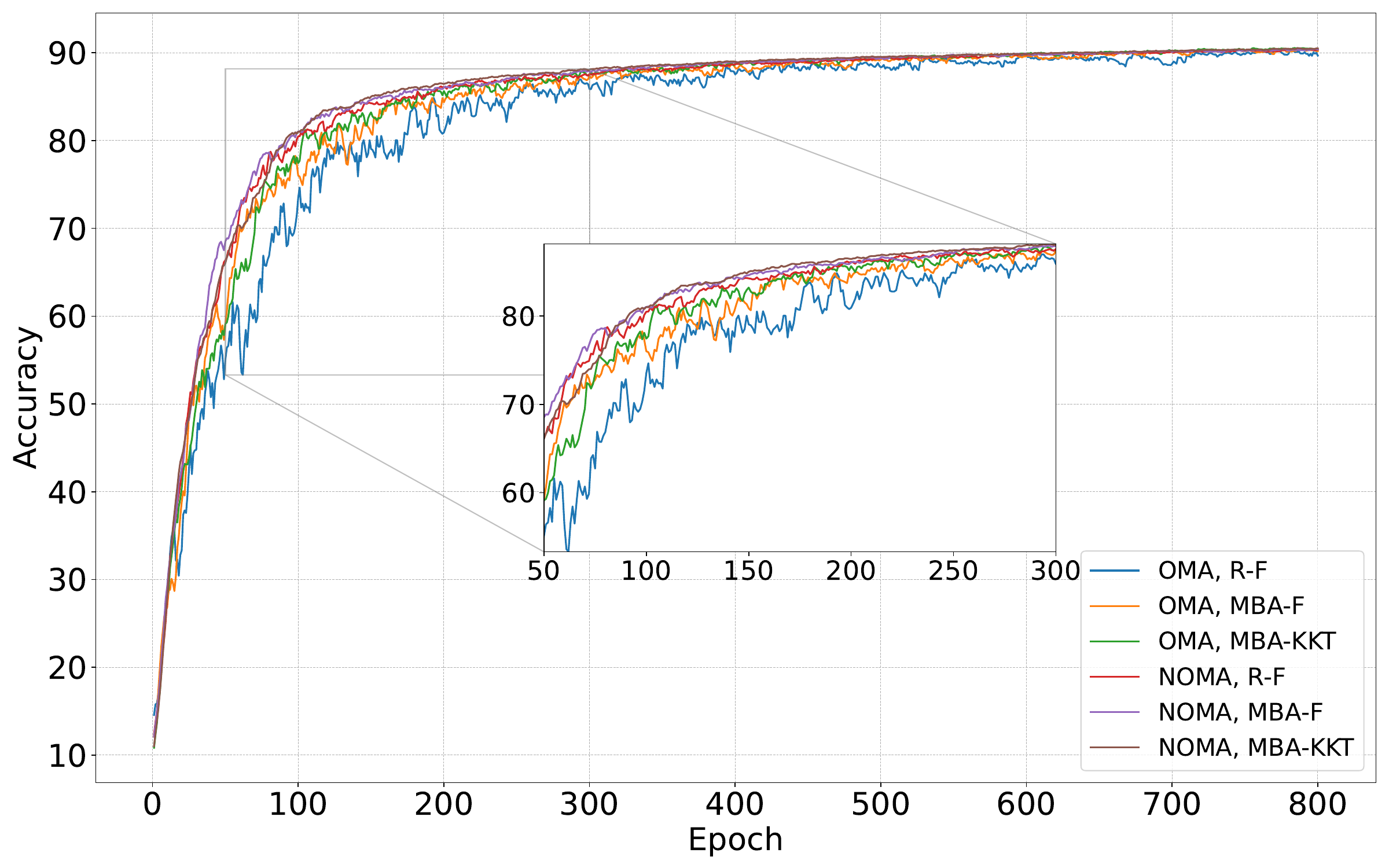}
    }
    
    \subfloat[Test accuracy on CIFAR-10.\label{Fig:output_cifar}]{%
        \includegraphics[width=0.9\linewidth]{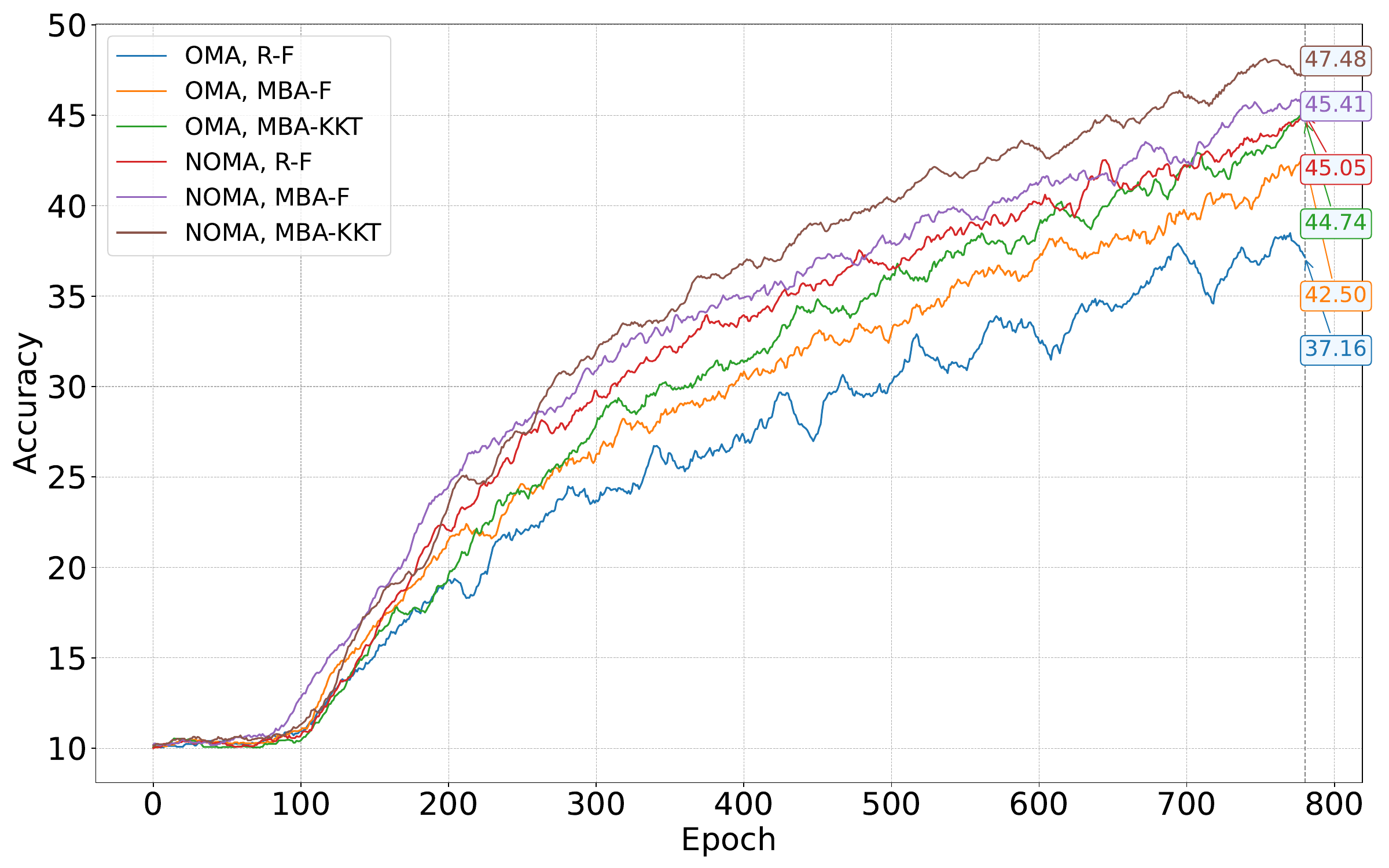}
    }
    \caption{The performance of the proposed framework on both datasets under different optimization strategies. $\alpha = 0.01$, $K = 10$, $T_{max}=6 s$.}
    \label{fig:FL_wireless}
\end{figure}
The model used to train MNIST is the neural network that begins by reshaping the input data from a 28x28 matrix to a flattened vector of 784 elements. This is followed by a fully connected layer, and then a dropout layer is employed for regularization with the Rectified Linear Unit (ReLU) activation function. Subsequently, another fully connected layer is applied before Softmax activation for the final classification output. 
For the model trained in the CIFAR-10 dataset, the architecture is similar to LeNet-5, which begins with a convolutional layer followed by ReLU activation function and a 2x2 max pooling. Subsequently, another convolution with 6 to 16 channels, is paired with ReLU and max pooling. After flattening, the data traverses through three fully connected layers, reducing dimensions progressively from the feature map size to 120, then 84, and finally 10. The output is processed via a Softmax activation for classification. The optimizer of training both datasets is stochastic gradient descent (SGD) for aligning the analysis in Section~\ref{section:system}, while 0.9 momentum is utilized to train the CIFAR-10 dataset for accelerating the convergence.
The model architectures used in the simulations are not state-of-the-art but are adequate for demonstrating the relative performance of the proposed methods in the paper.

Fig.~\ref{fig:FL_wireless} illustrates the performance comparisons between the proposed FL algorithm using MNIST and CIFAR-10 datasets under the extreme non-IID case similar in Fig.~\ref{Fig:alpha_001}. These comparisons are under different transmission strategies, including the random sub-channel allocation with fixed power allocation (R-F), matching-based sub-channel with fixed power allocation (MBA-F), and matching-based sub-channel with optimized power allocation (MBA-KKT), incorporating both in OMA and NOMA schemes. While the improvement in test accuracy for the MNIST dataset is moderate, it is evident that the proposed schemes, specifically MBA-KKT, can outperform on both datasets in terms of test accuracy and convergence rate. The improvement in performance is more notable in the CIFAR-10 dataset, indicating that the more challenging tasks benefit more from the proposed approaches. This improvement is attributed to the increased user participation during training, by optimizing through the proposed sub-channel and power allocations.
To ensure equitable comparisons, simulations carried out on the OMA scheme involve the same number of users as those in the NOMA scheme. For better visualization, the moving average technique is applied to the test accuracy data for the CIFAR-10 dataset. This process takes a set of data points and replaces each point with the average of the surrounding points. 

From comprehensive grid searches, it is notable to observe the effect of varying learning rates under the FL in wireless network settings. One of the observations is that the more users engage in training through optimal sub-channel and power allocation, the lesser the impact learning rate and batch size possess. It means that the optimization schemes could benefit the generalization performance and convergence rate, but also in fine-tuning processes. 
Additionally, it is not advisable to maintain a constant learning rate when an increasing number of users are contributing to the training. Intriguingly, as the number of users increases, the test accuracy also improves when the learning rate is raised, but only up to a certain point due to a larger batch size with a higher learning rate might lead to unstable training, potentially resulting the problems, i.e., overfitting or poor generalization. This observation contradicts the intuition that the learning rate should decrease as the batch size grows. 
This insight suggests that the learning rate could be proportionally increased with the batch size up to a certain limit. Exploring this relationship further could provide valuable guidance for the future design of advanced FL frameworks and optimization schemes, which may be included in future works.

\begin{figure}[!htbp] 
    \centering
    \includegraphics[width=0.95\linewidth]{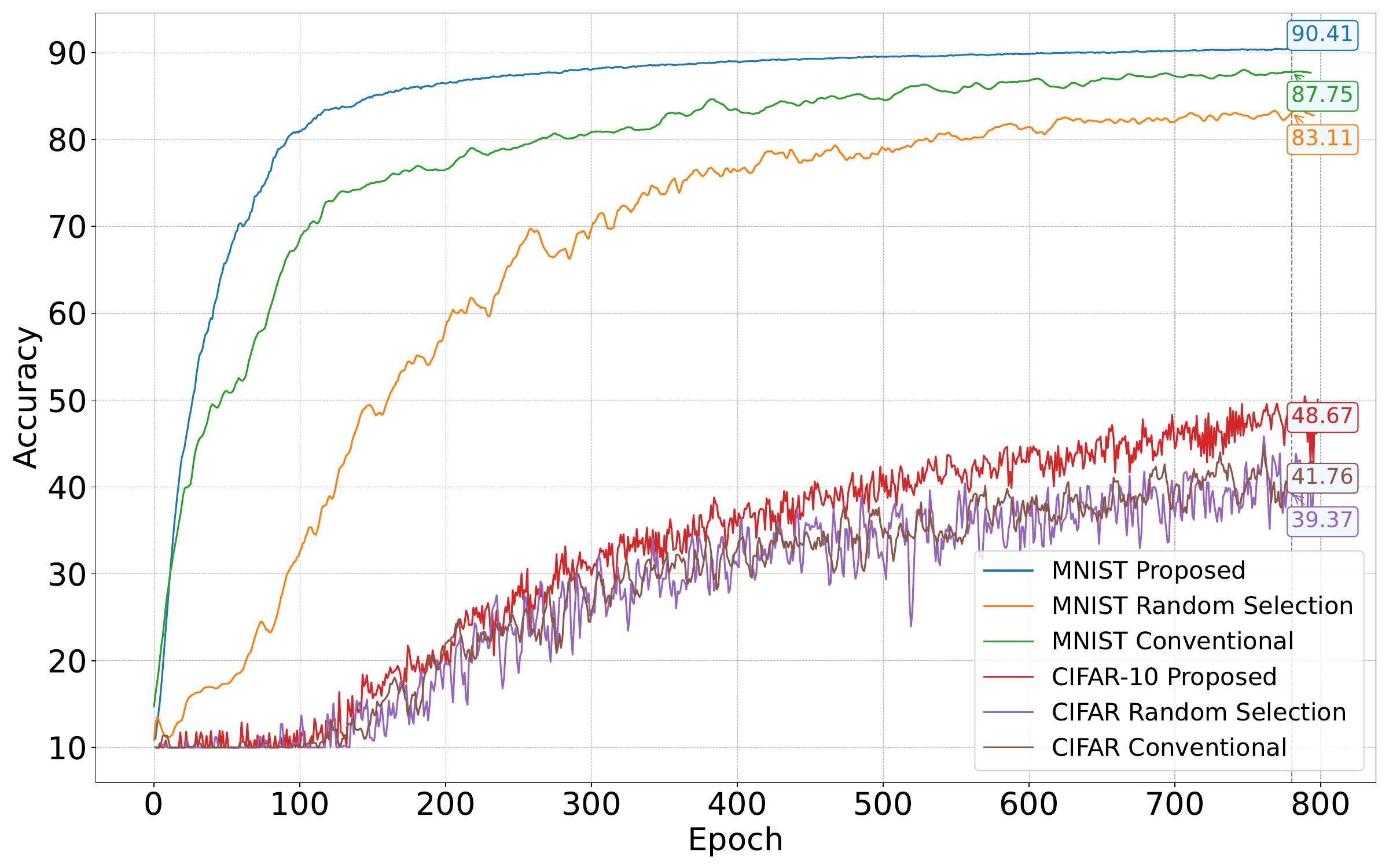}
    \caption{Comparisons between conventional FL, random selection, and the proposed method. $\alpha = 0.01$, $K = 10$, $T_{max}=6 s$.}
    \label{Fig:wireless_compare}
\end{figure}

The effectiveness of the proposed framework is evaluated using the MNIST and CIFAR-10 datasets, as shown in Fig.~\ref{Fig:wireless_compare}. The proposed framework can outperform both conventional FL and random selection approaches. In the latter approach, users are randomly assigned to clusters while keeping the same number of clusters as simulated in the proposed setup. It can be observed that neglecting the underlying distributions across the devices when employing random allocation can adversely affect performance. For example, accuracy drops from 87.75\% to 83.11\% in MNIST and from 41.76\% to 39.37\% in CIFAR-10. Moreover, the stability of training is affected compared to the conventional and the proposed framework. This instability can be attributed to the reduced number of users engaged in the specific group, negatively impacting both performance and stability, as previously demonstrated in Section~\ref{section:system}. Conversely, by carefully considering the underlying distributions, not only the generalization performance can be improved, but also the convergence rate could be beneficial as illustrated in Fig.~\ref{Fig:wireless_compare}.

\begin{figure}[t]
  \centering
  \subfloat[Cluster study on MNIST. $\alpha = 0.5$, $N=800$.]{
    \includegraphics[width=0.9\linewidth]{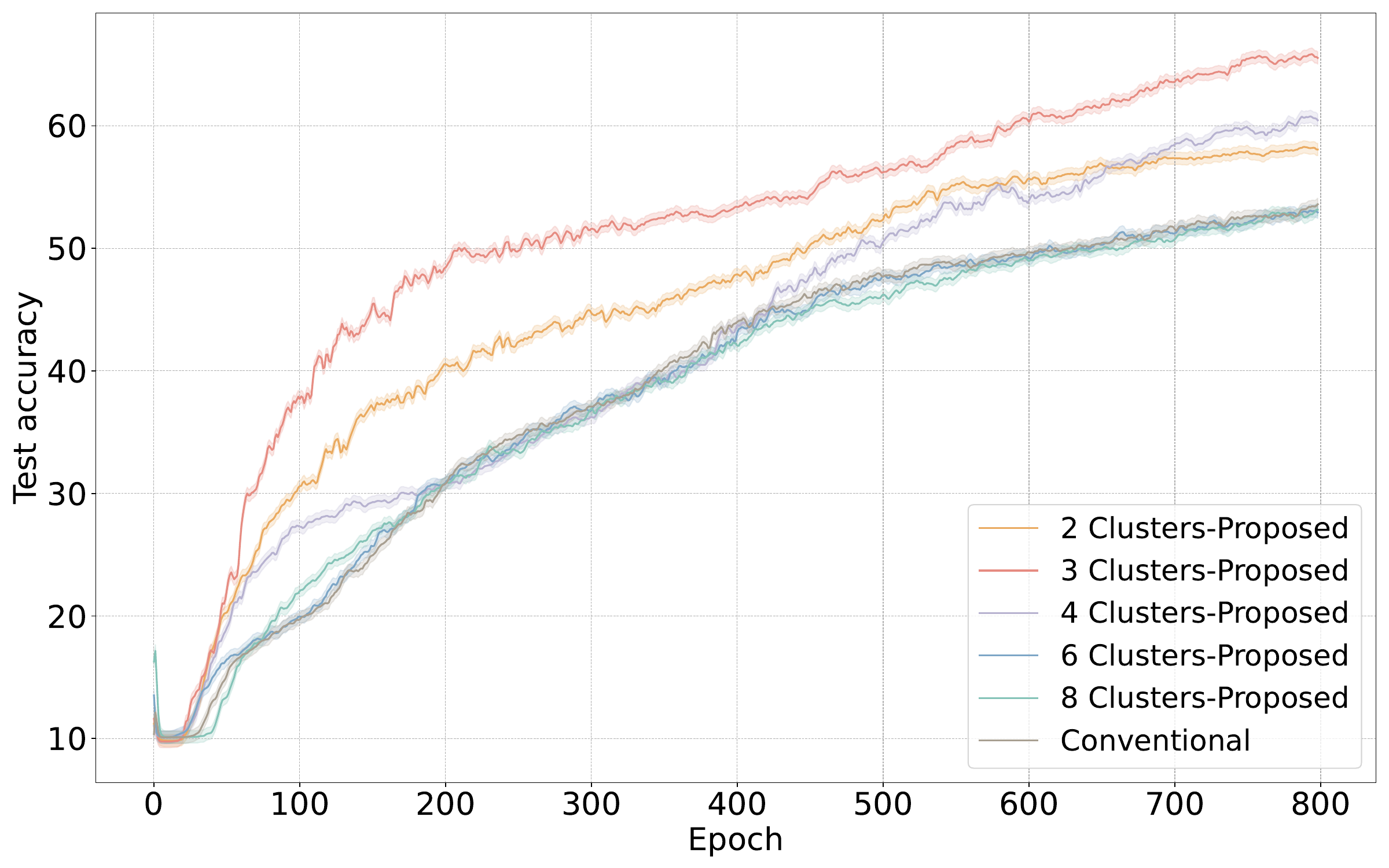}
    \label{fig:mnist_cluster}
  }
  \hfill
  \subfloat[Cluster study on CIFAR-10. $\alpha = 0.05$, $N=100$.]{
    \includegraphics[width=0.92\linewidth]{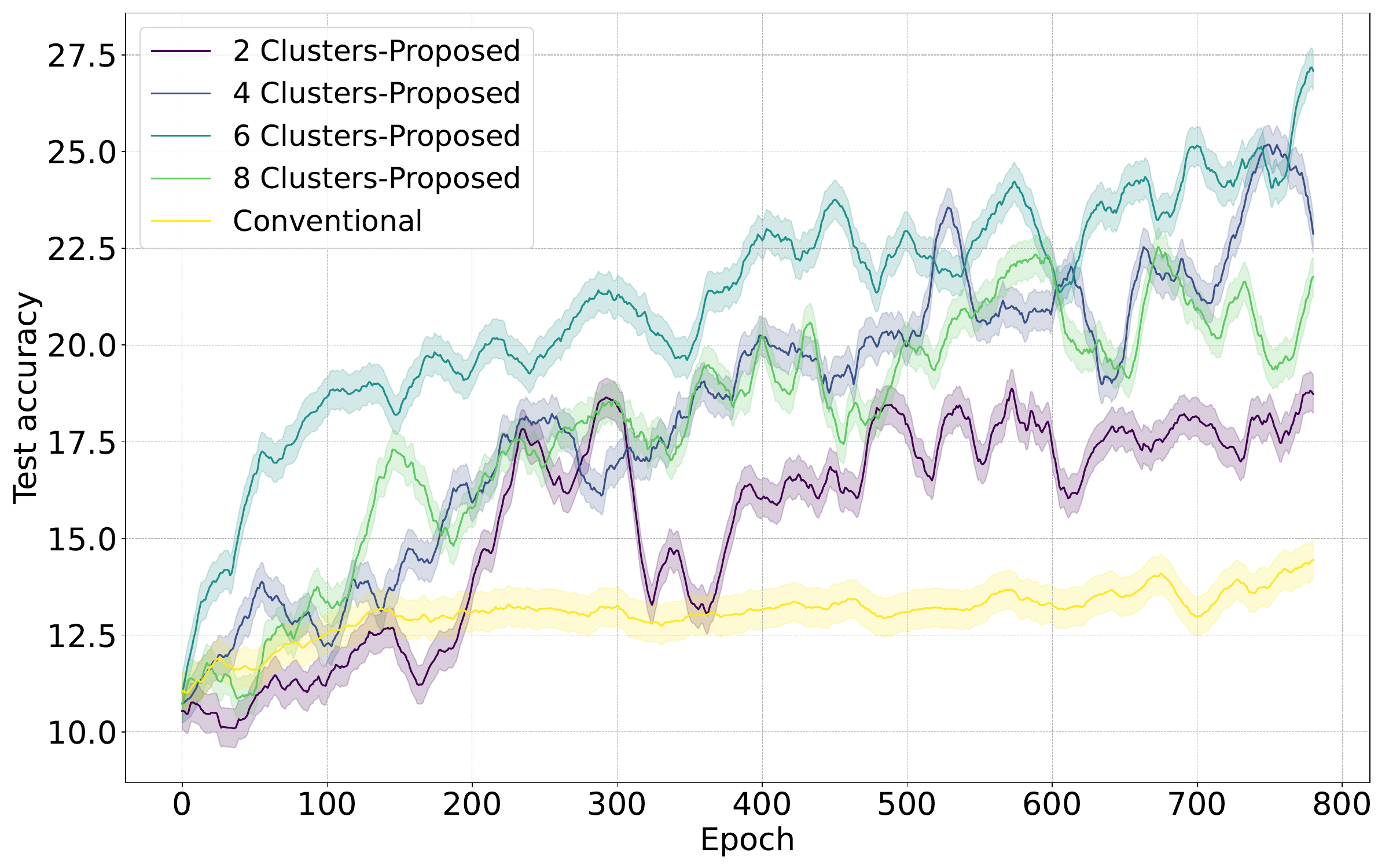}
    \label{fig:compare_cluster}
  }
  \caption{The impact of the different number of clusters on the performance in different datasets. $K=5$, $T_{max} = 6s$.}
  \label{fig:clusters}
\end{figure}

\begin{figure}[t]
    \centering
    {\includegraphics[width=0.9\linewidth]{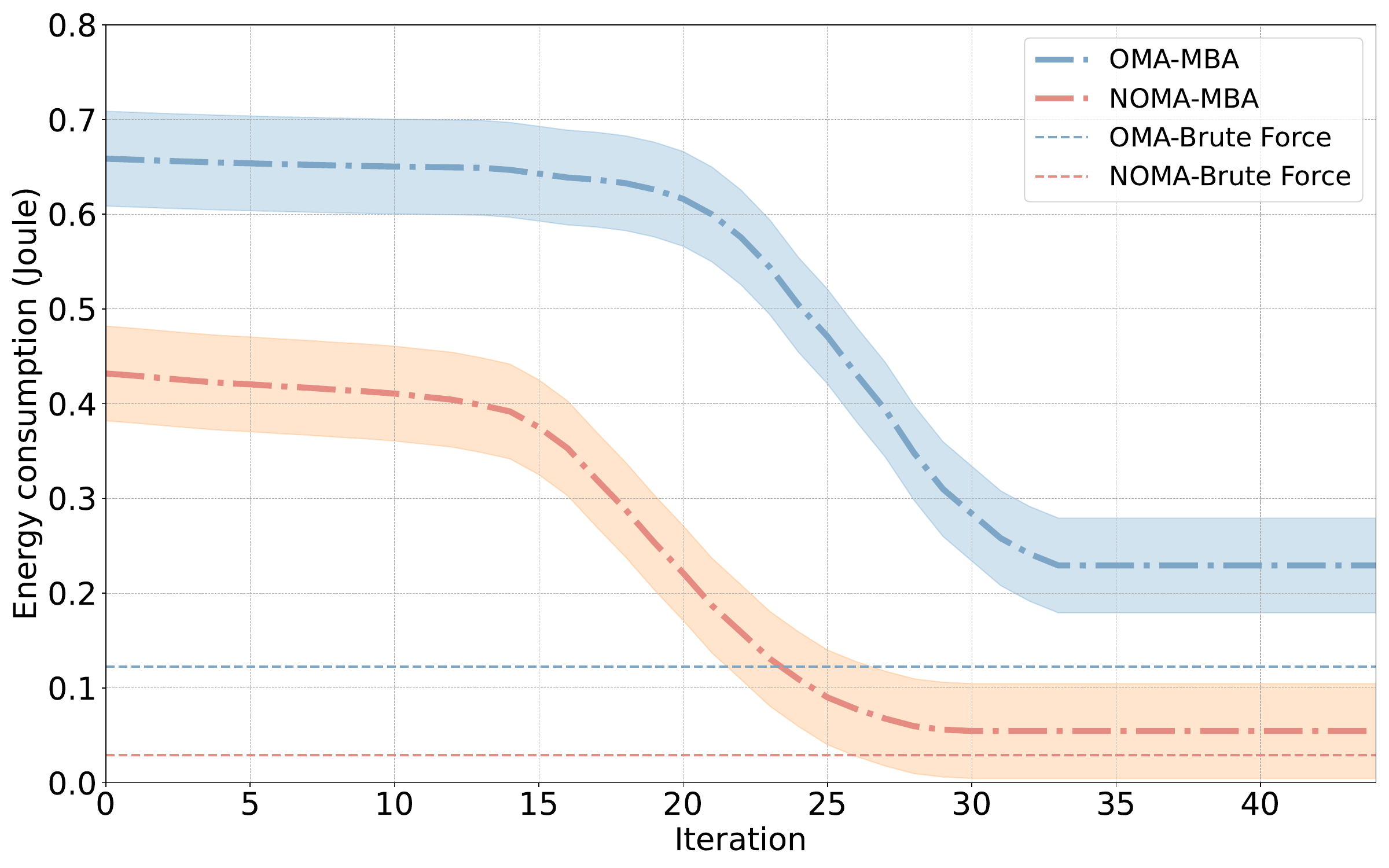}}
    \\
  \caption{The iterations of the matching-based algorithm. $K = 5$, $N = 10$, transmit time = $0.2s$.}
  \label{fig:energy}
\end{figure}
The impact of selecting the different number of clusters on the performance is studied in Fig.~\ref{fig:clusters}. To ensure balanced comparisons, a subset of users is selected to participate in the training process. Specifically, 10 out of 800 users and 10 out of 100 users are selected to join in the MNIST and CIFAR-10 datasets, respectively. 
The choice is motivated by the fact that the MNIST dataset typically has a less complex task for neural networks to learn compared to the CIFAR-10 dataset, prompting a larger $N$ value for a richer comparison in the clustering process. It can be observed that the proposed CFL can outperform the conventional FL. This is because the second term in Lemma 1 is minimized in the proposed method when the groups of users are carefully selected based on their underlying distributions, thereby enhancing the test accuracy as shown in Fig.~\ref{fig:clusters}. 
When the data distribution turns heterogeneous, i.e., $\alpha = 0.5$, the optimal number of clusters is determined to be 3, while when the distribution is more heterogeneous, i.e., $\alpha = 0.05$, the number of clusters is set to 6. 
Both the selected number of clusters on the proposed method perform the best, which demonstrates the effectiveness of the method demonstrated in~\ref{subsection:select_clusters}.
Additionally, both sub-figures illustrate that performance drops when an excessive number of clusters are chosen during training. Specifically, performance suffers when the number of clusters exceeds the optimal threshold. This further emphasizes the point discussed in subsection~\ref{subsection:select_clusters}, the number of clusters should not be brutally increased with the rise in the degree of heterogeneity. 

\vspace{-0.2cm}
\begin{figure}[!htbp] 
    \centering
{    \includegraphics[width=1\linewidth]{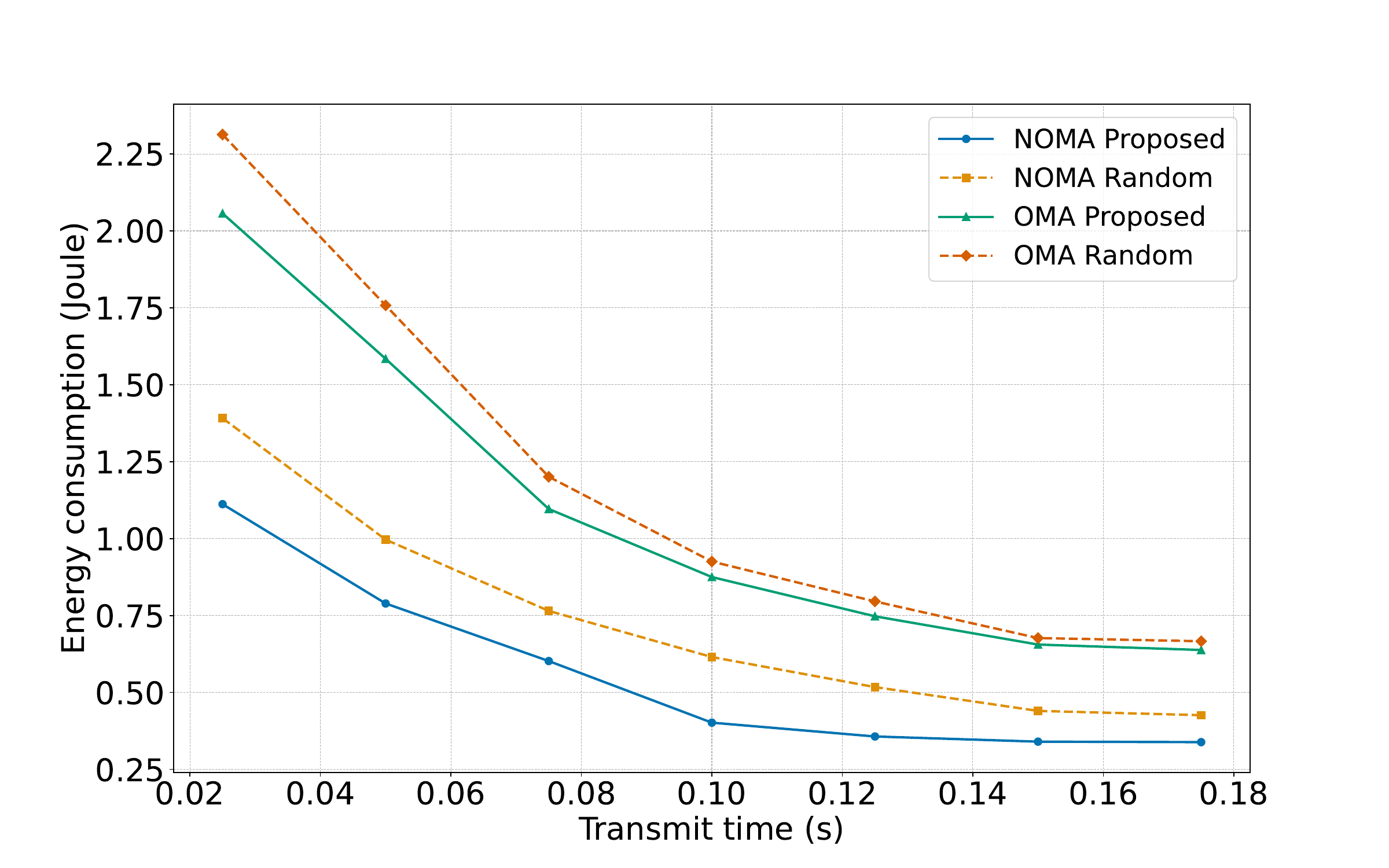}}
    \\
  \caption{The study on the energy consumption on transmit time. $K = 5$, $N = 10$.}
  \label{fig:energy_transmit}
\end{figure}
The study of iterations of the proposed sub-channel allocation algorithm using the solutions derived from KKT conditions under both OMA and NOMA schemes is shown in Fig.~\ref{fig:energy}, where the brute-force search is used to obtain the optimal solution. 
Observations from Fig.~\ref{fig:energy} indicate that the proposed algorithm reliably reaches a stable matching within roughly 30 iterations and obtains a performance that is approximately 80\% of the global optimum. Regarding computational efficiency, the proposed sub-channel allocation method is around 335 times faster than the exhaustive search-based strategy, as determined over 2000 simulations, averaging the execution times. The proposed method offers potential advantages, notably by greatly reducing the execution time when compared to the brute-force method and reducing energy consumption compared to the OMA scheme.

Finally, the impact of the transmit time on the energy consumption is studied as shown in Fig.~\ref{fig:energy_transmit}. It illustrates that permitting a greater transmit time can reduce energy use. This further confirms the conclusion that energy consumption is inversely related to transmit time as stated in Proposition 1. 
The simulation result demonstrates that employing the proposed matching-based algorithm with the closed-form solutions of KKT conditions outperforms random allocation in both OMA and NOMA schemes, illustrating that NOMA is more energy-efficient than the OMA.

\vspace{-0.1cm}
\section{Conclusions} \label{section:conclusion}

This paper has considered the integration of NOMA into the novel CFL framework by enabling the groups of devices to learn more specialized models through the clustering process, taking into account the time limitations and a finite number of sub-channels. Specifically, the server will learn the underlying distribution of the users, and gather the users with similar data distributions. The considered sub-channel and power allocation problem have been jointly optimized, and simulation results have been presented to demonstrate that the proposed FL framework can achieve drastic improvements over the FL baselines. Notably, integrating NOMA can further boost the performance of FL in terms of classification accuracy, by encouraging more users to participate during the training. A promising future direction is to explore the potential of active learning in the FL for labeling the data under the NOMA scheme. Furthermore, investigations of how NOMA can augment system throughput for real-time model transmission present an exciting direction for future research.

\section*{Appendix A:  Proof of Theorem~\ref{thm:core_motivation}}\label{Appendix_A}

In order to prove Theorem~\ref{thm:core_motivation}, a few definitions and notations are provided first. 
The inequality to state that with a probability of at least $1-\delta$ is introduced with the concept of McDiarmid’s inequality, providing a bound on the amount by which the function can deviate from its expectation. 
Note that the conventional Rademacher complexity loses its applicability in FL as the global loss function is the aggregate sum of the local objective functions, and therefore, the weighted Rademacher complexity is defined below.

\begin{defn}\label{defn:local_rademacher}
    Let $\mathcal{L}$ be the family of loss functions defined above, the empirical weighted Rademacher complexity defined for hypothesis space $\mathcal{H}$ based on Assumption 1 can be expressed as follows:
$$
\widehat{\mathcal{R}}_{\boldsymbol{\alpha}}(\mathcal{H})=\mathbb{E}_{\boldsymbol{\sigma}}\left[\sup _{\boldsymbol{w}\in \mathcal{H}} \sum_{i=1}^N \frac{A}{\beta_i a_i} \sum_{k=1}^{\beta_i} \sigma_{i,k} \ell_f\left(\boldsymbol{w}_{i,k}\right)\right],
$$
where $\sigma_{i,k} \in \{-1,+1\}$ are independent Rademacher variables, and $\ell_f\left(\boldsymbol{w}_{i,k}\right)$ is short for $\ell_f\left(\boldsymbol{w}, \boldsymbol{x}_{i,k},y_{i,k}\right)$. The weighted Rademacher complexity of $\mathcal{L}$ is $\mathcal{R}_{\boldsymbol{\alpha}}(\mathcal{H})=\mathbb{E}[\widehat{\mathcal{R}}_{\boldsymbol{\alpha}}(\mathcal{H})]$. The concept of Rademacher complexity and statistical learning theory refers to \cite{Rademacher_reference}.
\end{defn}

\begin{IEEEproof}
$\left|F_u\left(\widehat{\boldsymbol{w}}\right) - F_u\left({\boldsymbol{\widehat{w}}^*}\right)\right| $ can be upper bounded as

\begin{align}
\begin{split}
&\leq \left|F_u\left(\widehat{\boldsymbol{w}}\right) - F_u\left({\boldsymbol{\widehat{w}}^*}\right)\right| + \left|F\left(\widehat{\boldsymbol{w}}\right) - F\left({\boldsymbol{\widehat{w}}^*}\right)\right| \\
& \leq 2 \sup _{\boldsymbol{w} \in \mathcal{H}}\left(\left|F_u(\boldsymbol{w}) - F(\boldsymbol{w})\right|\right),
\end{split}
\end{align}
where the inequality holds because of the fact that individual differences between true and empirical risk will not exceed the maximum difference over all hypotheses in $\mathcal{H}$. 

Without loss of generality, denote $F'\left(\boldsymbol{w}\right)$ by the empirical loss function  by changing only one data point from the input space, and then the term can be further bounded as follows:
\begin{align}
 \begin{split} 
&\sup _{\boldsymbol{w} \in\mathcal{H}}\left|F_u(\boldsymbol{w}) - F'\left(\boldsymbol{w}\right)\right| - \sup _{\boldsymbol{w} \in\mathcal{H}}\left|F_u(\boldsymbol{w}) - F(\boldsymbol{w})\right| \\
&\leq \sup _{\boldsymbol{w} \in\mathcal{H}}\left|F_u(\boldsymbol{w}) - F'\left(\boldsymbol{w}\right) - F_u(\boldsymbol{w}) + F(\boldsymbol{w})\right| \\ 
& \leq \sup \left|F(\boldsymbol{w}) - F'\left(\boldsymbol{w}\right)\right|,
 \end{split}
\end{align}
Assuming the loss function is bounded by $b$,  then the term 
$\sup _{\boldsymbol{w} \in\mathcal{H}}\left|F_u(\boldsymbol{w}) - F'\left(\boldsymbol{w}\right)\right| - \sup _{\boldsymbol{w} \in\mathcal{H}}\left|F_u(\boldsymbol{w}) - F(\boldsymbol{w})\right| \leq \sum_{i\in \boldsymbol{S}}\frac{A b}{\beta_i a_i}$ holds due to the central tendency of each component in $\mathcal{MD}$ distribution, while changing one data point in input space will only make the function vary less than $\sum_{i\in \boldsymbol{S}}\frac{A b}{\beta_i a_i}$, and then McDiarmid's inequality can be used to further bound the term above following that with probability at least $1- \delta$: 
\begin{equation}
\begin{aligned}
&\sup _{\boldsymbol{w} \in\mathcal{H}}\left|F_u(\boldsymbol{w}) - F(\boldsymbol{w})\right| \leq \mathop{\mathbb{E}} \left[\sup _{\boldsymbol{w} \in\mathcal{H}}\left|F_u(\boldsymbol{w}) - F(\boldsymbol{w})\right|\right] + \Upsilon,
    \end{aligned}
\end{equation}
where $\Upsilon$ represents $b \sqrt{\sum _{i \in \boldsymbol{S}}\frac{A^2 \log(\frac{2}{\delta})}{2\beta_i^2 \alpha_i^2}}$. Now the term of $\mathop{\mathbb{E}} \left[\sup _{\boldsymbol{w} \in\mathcal{H}}\left|F_u(\boldsymbol{w})-F(\boldsymbol{w})\right|\right]$ can be further bounded by utilizing Jensen's inequality:
\begin{equation}
\begin{aligned}
    \begin{split}
        &\mathop{\mathbb{E}} \left[\sup _{\boldsymbol{w} \in\mathcal{H}}\left|F_u(\boldsymbol{w})-F(\boldsymbol{w})\right|\right] \\ 
        & = \mathop{\mathbb{E}} \left[\sup _{\boldsymbol{w} \in\mathcal{H}}\left|F_u(\boldsymbol{w})-F'(\boldsymbol{w})\right| - \sup _{\boldsymbol{w} \in\mathcal{H}}\left|F_u(\boldsymbol{w})-F(\boldsymbol{w})\right|\right] \\
        & \leq \mathop{\mathbb{E}} \left[\sup _{\boldsymbol{w} \in\mathcal{H}} \left|F_u(\boldsymbol{w})-F'(\boldsymbol{w}) - F_u(\boldsymbol{w})+F(\boldsymbol{w})\right| \right] \\
        & = \mathop{\mathbb{E}_{\boldsymbol{\sigma}}} \left[\sup _{\boldsymbol{w} \in\mathcal{H}} \sum_{i=1}^N \frac{A}{\beta_i a_i} \sum_{k=1}^{\beta_i} \sigma_{i,k} \left(\ell_f'(\boldsymbol{w}_{i,k})-\ell_f(\boldsymbol{w}_{i,k})\right)  \right] \\
        & \leq \mathop{\mathbb{E}_{\boldsymbol{\sigma}}} \left[\sup _{\boldsymbol{w} \in\mathcal{H}} \sum_{i=1}^N \frac{A}{\beta_i a_i} \sum_{k=1}^{\beta_i} \sigma_{i,k} \ell_f'(\boldsymbol{w}_{i,k}) \right] \\
        & \quad + \mathop{\mathbb{E}_{\boldsymbol{\sigma}}} \left[\sup _{\boldsymbol{w} \in\mathcal{H}} \sum_{i=1}^N \frac{A}{\beta_i a_i} \sum_{k=1}^{\beta_i} -\sigma_{i,k} \ell_f'(\boldsymbol{w}_{i,k}) \right] \\
        & = 2 \mathop{\mathbb{E}} [\widehat{\mathcal{R}}_{\boldsymbol{\alpha}}(\mathcal{H})],
    \end{split}
\end{aligned}
\end{equation}
where $\ell_f'(\boldsymbol{w}_{i,k})$ denotes the loss function by changing only one data point from the input space. The inequalities hold due to the fact that $-\sigma$ and $\sigma$ have the same distribution and the last inequality above follows Definition~\ref{defn:local_rademacher}. Then McDiarmid’s
inequality can be applied again. With probability at least $1 - \delta$, the following inequality can be obtained:
\begin{equation}
    \begin{aligned}
    \mathop{\mathbb{E}} [\widehat{\mathcal{R}}_{\boldsymbol{\alpha}}(\mathcal{H})] \leq \mathcal{R}_{\boldsymbol{\alpha}}(\mathcal{H}) + \Upsilon.
    \end{aligned}
\end{equation}
The derivations above lead to the final proof, in which the generalization gap can be rewritten as follows:
$$
\begin{aligned}
F_u\left(\widehat{\boldsymbol{w}}\right) - F_u\left({\boldsymbol{\widehat{w}}^*}\right) \leq 2 R_{\boldsymbol{\alpha}}(\mathcal{H}) + \sqrt{\sum _{i \in \boldsymbol{S}}\frac{ A^2 \log(\frac{2}{\delta})}{8 \beta_i^2 \alpha_i^2}},
\end{aligned}
$$
and the proof is completed.
\end{IEEEproof}

\newpage
\section*{Appendix B: Proof of Theorem~\ref{thm:convergence}}\label{AppendixB:convergence}
To prove Theorem 2, first expand the global loss function using the second-order Taylor expansion. Then the global loss function can be bounded as follows:
\begin{equation}
\begin{aligned}
& \mathbb{E} \left[F\left(\boldsymbol{w}^{\mathrm{t}}\right)\right] \\
\leq & \mathbb{E}\left[F\left(\boldsymbol{w}^{\mathrm{t}-1}\right)\right]+\frac{L}{2} \mathbb{E}\left[\left|-\eta \nabla F\left(\boldsymbol{w}^{\mathrm{t}-1}\right)\right|^2\right] \\
& +\mathbb{E}\left[\nabla F\left(\boldsymbol{w}^{\mathrm{t}-1}\right)^{\top}\left(-\eta \nabla F\left(\boldsymbol{w}^{\mathrm{t}-1}\right)\right)\right] \\
\leq & \mathbb{E}\left[F\left(\boldsymbol{w}^{\mathrm{t}-1}\right)\right]-\eta\left(1-\frac{\eta L}{2}\right)\left|\nabla F\left(\boldsymbol{w}^{\mathrm{t}-1}\right)\right|^2 \\
\end{aligned}
\end{equation}
With given that $\eta=\frac{1}{L}$,
$$
\begin{aligned}
\mathbb{E} \left[F\left(\boldsymbol{w}^{\mathrm{t}}\right)\right] \leq  \mathbb{E} \left[F\left(\boldsymbol{w}^{\mathrm{t}-1}\right)\right]-\frac{\eta}{2} \left|\nabla F\left(\boldsymbol{w}^{\mathrm{t}-1}\right)\right|^2 \\
\end{aligned}
$$
Denote vectors $\mathbf{a}=\left(\frac{\beta_1}{\sum_{i\in\boldsymbol{S}_{t-1}}\beta_i},\frac{\beta_2}{\sum_{i\in\boldsymbol{S}_{t-1}}\beta_i},\ldots,\frac{\beta_i}{\sum_{i\in\boldsymbol{S}_{t-1}}\beta_i}\right)$ and $\mathbf{b}=\left(\nabla f_1(\boldsymbol{w}^{t-1}),\nabla f_2(\boldsymbol{w}^{t-1}),\ldots,\nabla f_i(\boldsymbol{w}^{t-1})\right)$, respectively. 
 Then the Cauchy-Schwarz inequality can be further bound the term $| \nabla F\left(\boldsymbol{w}^{\mathrm{t}-1}\right)|^2$ by expressing it as the dot product of $\mathbf{a}$ and $\mathbf{b}$, i.e., $|\sum_{i\in\boldsymbol{S}_{t-1}}\frac{\beta_i}{\sum_{i\in\boldsymbol{S}_{t-1}}\beta_i}\nabla f_i(\boldsymbol{w}^{t-1})|^2 = \left|\mathbf{a}\cdot\mathbf{b}\right|^2$, which leads to:
$$
\begin{aligned}
&\left|\nabla F\left(\boldsymbol{w}^{\mathrm{t}-1}\right)\right|^2 \\
& \leq\bigg(\sum_{i\in\boldsymbol{S}_{t-1}}\frac{\beta_i^2}{\big(\sum_{i\in\boldsymbol{S}_{t-1}}\beta_i\big)^2}\bigg)\bigg(\sum_{i\in\boldsymbol{S}_{t-1}}|\nabla f_i(\boldsymbol{w}^{t-1})|^2\bigg), \\
\end{aligned}
$$
rearranging terms gives:
$$
\begin{aligned}
\left|\nabla F\left(\boldsymbol{w}^{\mathrm{t}-1}\right)\right|^2 \leq \frac{\sum_{i\in\mathcal{S}_{t-1}}\beta_i^2\sum_{i\in\mathcal{S}_{t-1}}|\nabla f_i(\boldsymbol{w}^{t-1})|^2}{\left(\sum_{i\in\mathcal{S}_{t-1}}\beta_i\right)^2}.
\end{aligned}
$$
Let the Assumption~\ref{assumption:bound_variance} holds, then the following inequality can be obtained using relaxed triangle inequality:
    \begin{equation} 
    \begin{aligned}
        & \mathbb E\left[\left| \nabla f_{i}(\boldsymbol{w})\right|^{2}\right] \\
        &= \mathbb E\left[\left| \nabla f_{i}(\boldsymbol{w}) - \nabla F(\boldsymbol{w}) + \nabla F(\boldsymbol{w})\right|^{2}\right] \\
        & \leq 2 \mathbb E\left[ \left| \nabla f_{i}(\boldsymbol{w}) - \nabla F(\boldsymbol{w}) \right|^2 \right] + 2 \mathbb E\left[\left| \nabla F(\boldsymbol{w})\right|^{2}\right] \\
        & \leq 2 G^2 + 2 \mathbb E\left[\left| \nabla F(\boldsymbol{w})\right|^{2}\right].
    \end{aligned}
    \end{equation}

Then the following inequality can be obtained as follows:
$$
\begin{aligned}
& \mathbb{E}\left[F\left(\boldsymbol{w}^{\mathrm{t}}\right)\right] \leq \mathbb{E}\left[F\left(\boldsymbol{w}^{\mathrm{t}-1}\right)\right] \\
& - \frac{\eta}{\left(\sum_{i \in \boldsymbol{S}_{t-1}} \beta_i\right)^2} \sum_{i \in \boldsymbol{S}_{t-1}} \beta_i^2 \left(G^2 + \mathbb E\left[\left| \nabla F(\boldsymbol{w}^{t-1})\right|^{2}\right]\right),
\end{aligned}
$$
subtract $F\left(\boldsymbol{w}^*\right)$ on both sides, then the final proof can be obtained as follows:
\begin{equation}
\begin{aligned}
&\mathbb{E}\left[F\left(\boldsymbol{w}^{\mathrm{t}}\right) - F\left(\boldsymbol{w}^*\right) \right] \leq \mathbb{E}\left[F\left(\boldsymbol{w}^{\mathrm{t}-1}\right)- F\left(\boldsymbol{w}^*\right)\right] \\
& - \frac{\eta}{\left(\sum_{i \in \boldsymbol{S}_{t-1}} \beta_i\right)^2} \sum_{i \in \boldsymbol{S}_{t-1}} \beta_i^2 \left(G^2 + \mathbb E\left[\left| \nabla F(\boldsymbol{w}^{t-1})\right|^{2}\right]\right),
\end{aligned}
\end{equation}
Then using $\mu$-PL inequality leads to the final derivation:
\begin{equation}
\begin{aligned}
&\mathbb{E}\left[F\left(\boldsymbol{w}^{\mathrm{t}}\right) - F\left(\boldsymbol{w}^*\right) \right] \\
\leq & \biggl(1 - \frac{\eta 2\mu}{\left(\sum_{i \in \boldsymbol{S}_{t-1}} \beta_i\right)^2} \sum_{i \in \boldsymbol{S}_{t-1}} \beta_i^2\biggl) \mathbb{E}\left[F\left(\boldsymbol{w}^{t-1}\right) - F\left(\boldsymbol{w}^*\right)\right] \\
& - \frac{\eta}{\left(\sum_{i \in \boldsymbol{S}_{t-1}} \beta_i\right)^2} \sum_{i \in \boldsymbol{S}_{t-1}} \beta_i^2 G^2,
\end{aligned}
\end{equation}

and the proof is completed.

\bibliographystyle{IEEEtran}
\bibliography{FL_reference}

\begin{thebibliography}{10}
\providecommand{\url}[1]{#1}
\csname url@samestyle\endcsname
\providecommand{\newblock}{\relax}
\providecommand{\bibinfo}[2]{#2}
\providecommand{\BIBentrySTDinterwordspacing}{\spaceskip=0pt\relax}
\providecommand{\BIBentryALTinterwordstretchfactor}{4}
\providecommand{\BIBentryALTinterwordspacing}{\spaceskip=\fontdimen2\font plus
\BIBentryALTinterwordstretchfactor\fontdimen3\font minus \fontdimen4\font\relax}
\providecommand{\BIBforeignlanguage}[2]{{%
\expandafter\ifx\csname l@#1\endcsname\relax
\typeout{** WARNING: IEEEtran.bst: No hyphenation pattern has been}%
\typeout{** loaded for the language `#1'. Using the pattern for}%
\typeout{** the default language instead.}%
\else
\language=\csname l@#1\endcsname
\fi
#2}}
\providecommand{\BIBdecl}{\relax}
\BIBdecl

\bibitem{VTC_Lin}
Y.~Lin, K.~Wang, and Z.~Ding, ``Sub-channel assignment and power allocation in {NOMA}-enhanced federated learning networks,'' \emph{submitted to 2024 {IEEE} 99th Vehicular Technology Conference (VTC Spring)}, 2023.

\bibitem{Cisco}
\BIBentryALTinterwordspacing
Cisco. Annual internet report (2018–2023) white paper. [Online]. Available: \url{https://www.cisco.com/c/en/us/solutions/collateral/executive-perspectives/annual-internet-report/white-paper-c11-741490.html}
\BIBentrySTDinterwordspacing

\bibitem{Y_Liu_FL_literature}
Y.~Liu, X.~Yuan, Z.~Xiong, J.~Kang, X.~Wang, and D.~Niyato, ``Federated learning for {6G} communications: Challenges, methods, and future directions,'' \emph{China Commun.}, vol.~17, no.~9, pp. 105--118, Sep. 2020.

\bibitem{FL_NOMA_6G}
P.~S. Bouzinis \emph{et~al.}, ``Wireless federated learning for {6G} networks—part ii: The compute-then-transmit {NOMA} paradigm,'' \emph{IEEE Commun. Lett.}, vol.~26, no.~1, pp. 8--12, Oct. 2022.

\bibitem{FL_Wireless_Qin}
Z.~Qin \emph{et~al.}, ``Federated learning and wireless communications,'' \emph{IEEE Wireless Commun.}, vol.~28, no.~5, pp. 134--140, Oct. 2021.

\bibitem{NOMA_user_association}
K.~Wang \emph{et~al.}, ``User association and power allocation for multi-cell non-orthogonal multiple access networks,'' \emph{IEEE Trans. Wireless Commun.}, vol.~18, no.~11, pp. 5284--5298, Nov. 2019.

\bibitem{NOMA_MEC_Ding}
Z.~Ding, P.~Fan, and H.~V. Poor, ``Impact of non-orthogonal multiple access on the offloading of mobile edge computing,'' \emph{IEEE Trans. Commun.}, vol.~67, no.~1, pp. 375--390, Jan. 2019.

\bibitem{FL_original}
\BIBentryALTinterwordspacing
H.~B. McMahan, E.~Moore, D.~Ramage, S.~Hampson, and B.~A.~y. Arcas, ``Communication-efficient learning of deep networks from decentralized data,'' 2016. [Online]. Available: \url{https://arxiv.org/abs/1602.05629}
\BIBentrySTDinterwordspacing

\bibitem{FL_literature_1}
\BIBentryALTinterwordspacing
P.~Kairouz and H.~B.~M. \textit{et al}, ``Advances and open problems in federated learning,'' 2021. [Online]. Available: \url{https://arxiv.org/abs/1912.04977}
\BIBentrySTDinterwordspacing

\bibitem{FL_with_non_iid}
\BIBentryALTinterwordspacing
Z.~Yue \emph{et~al.}, ``Federated learning with non-{IID} data,'' 2018. [Online]. Available: \url{https://arxiv.org/abs/1806.00582}
\BIBentrySTDinterwordspacing

\bibitem{scaffold_FL}
\BIBentryALTinterwordspacing
S.~P. Karimireddy, S.~Kale, M.~Mohri, S.~J. Reddi, S.~U. Stich, and A.~T. Suresh, ``{SCAFFOLD:} stochastic controlled averaging for on-device federated learning,'' \emph{CoRR}, vol. abs/1910.06378, 2019. [Online]. Available: \url{http://arxiv.org/abs/1910.06378}
\BIBentrySTDinterwordspacing

\bibitem{CFL_model_agnostic}
F.~Sattler, K.-R. Müller, and W.~Samek, ``Clustered federated learning: Model-agnostic distributed multitask optimization under privacy constraints,'' \emph{IEEE Trans. Neural Netw. Learn. Syst.}, vol.~32, no.~8, pp. 3710--3722, Aug. 2021.

\bibitem{HypCluster}
\BIBentryALTinterwordspacing
Y.~Mansour \emph{et~al.}, ``Three approaches for personalization with applications to federated learning,'' \emph{CoRR}, vol. abs/2002.10619, 2020. [Online]. Available: \url{https://arxiv.org/abs/2002.10619}
\BIBentrySTDinterwordspacing

\bibitem{FL_HC}
\BIBentryALTinterwordspacing
C.~Briggs, Z.~Fan, and P.~Andras, ``Federated learning with hierarchical clustering of local updates to improve training on non-{IID} data,'' \emph{CoRR}, vol. abs/2004.11791, 2020. [Online]. Available: \url{https://arxiv.org/abs/2004.11791}
\BIBentrySTDinterwordspacing

\bibitem{FedMe}
\BIBentryALTinterwordspacing
K.~Matsuda, Y.~Sasaki, C.~Xiao, and M.~Onizuka, ``Fedme: Federated learning via model exchange,'' \emph{CoRR}, vol. abs/2110.07868, 2021. [Online]. Available: \url{https://arxiv.org/abs/2110.07868}
\BIBentrySTDinterwordspacing

\bibitem{kaidi2023fl2}
K.~Wang, Z.~Ding, D.~K.~C. So, and Z.~Ding, ``Age-of-information minimization in federated learning based networks with {Non-IID} dataset,'' \emph{submitted to {IEEE} Trans. Wireless Commun.}, 2023.

\bibitem{FL_fading_channels}
M.~M. Amiri and D.~Gündüz, ``Federated learning over wireless fading channels,'' \emph{IEEE Trans. Wireless Commun.}, vol.~19, no.~5, pp. 3546--3557, May 2020.

\bibitem{joint_communications_FL}
M.~Chen \emph{et~al.}, ``A joint learning and communications framework for federated learning over wireless networks,'' \emph{IEEE Trans. Wireless Commun.}, vol.~20, no.~1, pp. 269--283, Jan. 2021.

\bibitem{kaidi_AOI_FL}
\BIBentryALTinterwordspacing
K.~Wang, Y.~Ma, M.~B. Mashhadi, C.~H. Foh, R.~Tafazolli, and Z.~Ding, ``Age of information in federated learning over wireless networks,'' 2022. [Online]. Available: \url{https://arxiv.org/abs/2209.06623}
\BIBentrySTDinterwordspacing

\bibitem{Contribution_based_WFL}
S.~R. Pandey, L.~D. Nguyen, and P.~Popovski, ``A contribution-based device selection scheme in federated learning,'' \emph{IEEE Commun. Lett.}, vol.~26, no.~9, pp. 2057--2061, Sept. 2022.

\bibitem{Latency_constrained_WFL}
W.~Shi \emph{et~al.}, ``Joint device scheduling and resource allocation for latency constrained wireless federated learning,'' \emph{IEEE Trans. Wireless Commun.}, vol.~20, no.~1, pp. 453--467, 2021.

\bibitem{FL_nanyang}
Y.-J. Liu \emph{et~al.}, ``Ensemble distillation based adaptive quantization for supporting federated learning in wireless networks,'' \emph{IEEE Trans. Wireless Commun.}, vol.~22, no.~6, pp. 4013--4027, Jun. 2023.

\bibitem{NOMA_survey_ding}
Z.~Ding, L.~Lv, F.~Fang, O.~A. Dobre, G.~K. Karagiannidis, N.~Al-Dhahir, R.~Schober, and H.~V. Poor, ``A state-of-the-art survey on reconfigurable intelligent surface-assisted non-orthogonal multiple access networks,'' \emph{Proc. IEEE}, vol. 110, no.~9, pp. 1358--1379, Sept. 2022.

\bibitem{NOMA_survey_fang}
Q.-V. Pham, F.~Fang, V.~N. Ha, M.~J. Piran, M.~Le, L.~B. Le, W.-J. Hwang, and Z.~Ding, ``A survey of multi-access edge computing in {5G} and beyond: Fundamentals, technology integration, and state-of-the-art,'' \emph{IEEE Access}, vol.~8, pp. 116\,974--117\,017, Jun. 2020.

\bibitem{ZhangSWIPT}
R.~Zhang, K.~Xiong, Y.~Lu, D.~W.~K. Ng, P.~Fan, and K.~B. Letaief, ``Swipt-enabled cell-free massive {MIMO-NOMA} networks: A machine learning-based approach,'' \emph{IEEE Trans. Wireless Commun.}, 2023.

\bibitem{Lin_unsupervised}
Y.~Lin, K.~Wang, and Z.~Ding, ``Unsupervised machine learning-based user clustering in {THz-NOMA} systems,'' \emph{IEEE Wireless Commun. Lett.}, vol.~12, no.~7, pp. 1130--1134, Jul. 2023.

\bibitem{Wenqi_NOMA}
W.~Huang and Z.~Ding, ``New insight for multi-user hybrid {NOMA} offloading strategies in mec networks,'' \emph{IEEE Trans. Veh. Technol.}, pp. 1--6, 2023.

\bibitem{Qin_NOMA}
Z.~Qin \emph{et~al.}, ``User association and resource allocation in unified {NOMA} enabled heterogeneous ultra dense networks,'' \emph{IEEE Commun. Mag.}, vol.~56, no.~6, pp. 86--92, Jun. 2018.

\bibitem{Adaptive_FL_NOMA}
H.~Sun, X.~Ma, and R.~Q. Hu, ``Adaptive federated learning with gradient compression in uplink {NOMA},'' \emph{IEEE Trans. Veh. Technol.}, vol.~69, no.~12, pp. 16\,325--16\,329, Dec. 2020.

\bibitem{NOMA_FL_cost}
Y.~Wu \emph{et~al.}, ``Non-orthogonal multiple access assisted federated learning via wireless power transfer: A cost-efficient approach,'' \emph{IEEE Trans. Commun.}, vol.~70, no.~4, pp. 2853--2869, Apr. 2022.

\bibitem{wanli_FL}
W.~Ni, Y.~Liu, Y.~C. Eldar, Z.~Yang, and H.~Tian, ``Enabling ubiquitous non-orthogonal multiple access and pervasive federated learning via star-ris,'' in \emph{2021 IEEE Global Communications Conference (GLOBECOM)}, 2021, pp. 1--6.

\bibitem{RIS_FL_NOMA}
R.~Zhong \emph{et~al.}, ``Mobile reconfigurable intelligent surfaces for {NOMA} networks: Federated learning approaches,'' \emph{IEEE Trans. Wireless Commun.}, vol.~21, no.~11, pp. 10\,020--10\,034, Nov. 2022.

\bibitem{Scheduling_allocation_FL}
X.~Ma, H.~Sun, and R.~Q. Hu, ``Scheduling policy and power allocation for federated learning in {NOMA}-based {MEC},'' in \emph{Proc. IEEE Global Commun. Conf.}, Dec. 2020, pp. 1--7.

\bibitem{Privacy_NOMA_FL}
D.~Liu and O.~Simeone, ``Privacy for free: Wireless federated learning via uncoded transmission with adaptive power control,'' \emph{IEEE J. Sel. Areas Commun.}, vol.~39, no.~1, pp. 170--185, Jan. 2021.

\bibitem{hybrid_NOMA}
Z.~Ding, D.~Xu, R.~Schober, and H.~V. Poor, ``Hybrid {NOMA} offloading in multi-user {MEC} networks,'' \emph{IEEE Trans. Wireless Commun.}, vol.~21, no.~7, pp. 5377--5391, Jul. 2022.

\bibitem{spectral}
\BIBentryALTinterwordspacing
A.~Ng, M.~Jordan, and Y.~Weiss, ``On spectral clustering: Analysis and an algorithm,'' in \emph{Adv. Neural Inf. Process}, vol.~14.\hskip 1em plus 0.5em minus 0.4em\relax MIT Press, 2001. [Online]. Available: \url{https://proceedings.neurips.cc/paper_files/paper/2001/file/801272ee79cfde7fa5960571fee36b9b-Paper.pdf}
\BIBentrySTDinterwordspacing

\bibitem{spectral_tutorial}
\BIBentryALTinterwordspacing
U.~von Luxburg, ``A tutorial on spectral clustering,'' \emph{CoRR}, vol. abs/0711.0189, 2007. [Online]. Available: \url{http://arxiv.org/abs/0711.0189}
\BIBentrySTDinterwordspacing

\bibitem{K_means_bound}
\BIBentryALTinterwordspacing
S.~Li and Y.~Liu, ``Sharper generalization bounds for clustering,'' in \emph{Proceedings of the 38th International Conference on Machine Learning}, ser. Proceedings of Machine Learning Research, vol. 139.\hskip 1em plus 0.5em minus 0.4em\relax PMLR, Jul. 2021, pp. 6392--6402. [Online]. Available: \url{https://proceedings.mlr.press/v139/li21k.html}
\BIBentrySTDinterwordspacing

\bibitem{on_learning_operator}
\BIBentryALTinterwordspacing
L.~Rosasco, M.~Belkin, and E.~D. Vito, ``On learning with integral operators,'' \emph{J Mach Learn Res.}, vol.~11, no.~30, pp. 905--934, 2010. [Online]. Available: \url{http://jmlr.org/papers/v11/rosasco10a.html}
\BIBentrySTDinterwordspacing

\bibitem{select_k_spectal}
\BIBentryALTinterwordspacing
C.~R. John, D.~Watson, M.~R. Barnes, C.~Pitzalis, and M.~J. Lewis, ``{Spectrum: fast density-aware spectral clustering for single and multi-omic data},'' \emph{Bioinformatics}, vol.~36, no.~4, pp. 1159--1166, 09 2019. [Online]. Available: \url{https://doi.org/10.1093/bioinformatics/btz704}
\BIBentrySTDinterwordspacing

\bibitem{Rademacher_reference}
\BIBentryALTinterwordspacing
O.~Bousquet, S.~Boucheron, and G.~Lugosi, \emph{Introduction to Statistical Learning Theory}.\hskip 1em plus 0.5em minus 0.4em\relax Berlin, Heidelberg: Springer Berlin Heidelberg, 2004, pp. 169--207. [Online]. Available: \url{https://doi.org/10.1007/978-3-540-28650-9_8}
\BIBentrySTDinterwordspacing

\end{thebibliography}
\end{document}